\documentclass[12pt,aps,prd,nofootinbib]{revtex4-2}

\usepackage[T1]{fontenc}
\usepackage[utf8]{inputenc}
\usepackage{lmodern}
\usepackage{amsmath,amssymb,bm}
\usepackage{graphicx}
\graphicspath{{./figures/}}
\usepackage{xcolor}
\usepackage[colorlinks=true,linkcolor=blue,citecolor=blue,urlcolor=blue]{hyperref}
\usepackage{float}
\usepackage{placeins}
\usepackage{subcaption}
\usepackage{booktabs}
\newcommand{\BR}{\operatorname{BR}}

\begin{document}

\title{Flavor-Violating Higgs and Top Decays in the Type 1B Flavorful Two-Higgs-Doublet Model with a Twist}

\author{Ti-Bin Hou$^{1,2,3}$}
\email{3078849770@qq.com}

\author{Zheng Li$^{1,2,3}$}

\author{Yu-Ju Peng$^{1,2,3}$}

\author{Jin-Lei Yang$^{1,2,3}$}
\email{jlyang@hbu.edu.cn}

\author{Hao-Ran Ma$^{1,2,3}$}

\author{Tai-Fu Feng$^{1,2,3,4}$}
\email{fengtf@hbu.edu.cn}

\affiliation{$^1$Department of Physics, Hebei University, Baoding 071002, China}
\affiliation{$^2$Hebei Key Laboratory of High-precision Computation and Application of Quantum Field Theory, Baoding 071002, China}
\affiliation{$^3$Hebei Research Center of the Basic Discipline for Computational Physics, Baoding 071002, China}
\affiliation{$^4$Department of Physics, Chongqing University, Chongqing 401331, China}

\begin{abstract}
\begin{center}
\textbf{Abstract}
\end{center}
\hspace*{1em}We study flavor-violating decays of the SM-like Higgs boson and rare top decays in the Type 1B realization of the flavorful two-Higgs-doublet model with a twist, focusing on $h\to sb$, $h\to db$, $h\to uc$, $h\to\mu\tau$, $h\to e\tau$, $h\to e\mu$, $t\to ch$, and $t\to uh$. We present the relevant amplitude and branching-ratio formulae, include one-loop vertex corrections, and impose constraints from Higgs signal-strength measurements, electroweak precision tests, meson mixing, $B\to X_s\gamma$, direct HFV searches and CLFV constraints, including $\mu-e$ conversion. The surviving parameter space favors $\BR(h\to sb)\gg\BR(h\to db)$, $\BR(t\to ch)\gg\BR(t\to uh)$, and $\BR(h\to\mu\tau)\gg\BR(h\to e\tau)\sim\BR(h\to e\mu)$ in the lepton sector. One-loop effects usually preserve these hierarchies, but can modify individual channels through different mechanisms: sizable destructive tree--loop interference can reduce $h\to sb$ enough to yield rare $\BR_{\mathrm{1L}}(h\to db)>\BR_{\mathrm{1L}}(h\to sb)$ points, whereas the large relative corrections found in rare top decays mainly reflect strong SM-like alignment suppressing the tree-level amplitudes.
\end{abstract}

\maketitle

\clearpage

\section{Introduction}

The discovery of the Higgs boson by the ATLAS and CMS Collaborations completed the particle content of the Standard Model (SM), leaving the observed pattern of fermion masses and flavor mixing as an open question \cite{ATLAS:2012yve,CMS:2012qbp}. Higgs interactions therefore remain a direct probe of the flavor puzzle. In a generic two-Higgs-doublet model (2HDM), the presence of several Yukawa matrices leads to tree-level flavor-changing neutral currents (FCNCs). The traditional solution is natural flavor conservation, in which each fermion species couples to only one scalar doublet \cite{Glashow:1976nt,Paschos:1976ay}. Yukawa alignment provides another well-known way to suppress scalar FCNCs \cite{Pich:2009sp}. These ideas underlie much of the conventional 2HDM classification and phenomenology \cite{Branco:2011iw}.

The absence of large observed FCNCs is compatible with a structured scalar flavor sector. Controlled scalar flavor violation can arise if the non-standard Yukawa couplings are correlated with fermion masses and mixing. Early examples include the Cheng--Sher ansatz and top-specific scalar flavor violation \cite{Cheng:1987rs,Atwood:1996vj,Hou:1991un}. Branco--Grimus--Lavoura constructions provide another important class in which scalar FCNC couplings are controlled by CKM matrix elements \cite{Branco:1996bq,Botella:2014ska,Botella:2015hoa}. These examples motivate extended Higgs sectors with structured flavor violation.

Flavorful two-Higgs-doublet models follow this philosophy. The flavor-locked flavorful 2HDM showed that characteristic Higgs signatures can be compatible with flavor data when the Yukawa sector is organized by a controlled flavor structure \cite{Altmannshofer:2017uvs}. The flavorful two-Higgs-doublet model with a twist introduced several twisted Yukawa realizations, including the Type 1B case studied here \cite{Altmannshofer:2018bch}. The Type 1B F2HDM has a non-standard flavor structure, approximate protection of first--second generation transitions, and in the implementation considered here a CKM matrix generated in the down sector, distinguishing it from the ordinary natural-flavor-conserving Type-I 2HDM.

The experimental situation makes Higgs and top flavor violation especially relevant. Direct LHC searches for charged-lepton-flavor-violating Higgs decays have reached sub-percent sensitivity in $h\to\mu\tau$ and $h\to e\tau$, while $h\to e\mu$ is constrained even more strongly \cite{CMS:2021rsq,ATLAS:2023mvd,CMS:2023xpx}. Charged-lepton flavor violation in heavy-particle decays, including Higgs, $Z$, and top decays, is a broad probe of new flavor dynamics \cite{Altmannshofer:2022bza}. Top-Higgs FCNC interactions are likewise well motivated because SM top FCNC branching fractions are extremely small \cite{Aguilar-Saavedra:2004mfd}. Top FCNC decays have also been analyzed in several \(U(1)\)-extended BSM frameworks, including flavor-dependent \(U(1)\) and \(B-L\)-type models \cite{Ge:2024esr,Yang:2018utw}. LHC searches for $t\to Hq$ in $H\to b\bar b$, $H\to\gamma\gamma$, and multi-lepton final states now probe branching fractions at the per-mille level or below \cite{CMS:2022vso,ATLAS:2023xpr,ATLAS:2024topHml}.

On the theory side, Higgs flavor violation has been studied in effective-coupling approaches, general 2HDMs, aligned 2HDMs, and flavorful Higgs frameworks \cite{Blankenburg:2012ex,Harnik:2012pb,Kopp:2014rva,Buschmann:2016uzg,Herrero-Garcia:2019mcy,Abbas:2015cua}. Collider signatures of flavorful Higgs bosons can differ substantially from those of natural-flavor-conserving 2HDMs \cite{Altmannshofer:2016zrn}. Rare top decays have also been proposed as direct probes of flavorful Higgs sectors \cite{Altmannshofer:2019ogm}. Related flavor-dependent Higgs-sector studies have investigated rare Higgs decays, top FCNC channels, charged-lepton-flavor violation and electroweak precision constraints in \(U(1)_F\)-type models \cite{Cao:2025bfb}. These studies provide the context for a joint numerical analysis of Higgs and top flavor violation in a Type 1B F2HDM.

In this work we analyze the channels $h\to sb$, $h\to db$, $h\to uc$, $h\to\mu\tau$, $h\to e\tau$, $h\to e\mu$, $t\to ch$, and $t\to uh$, using the charge-conjugate convention specified below. The final sample is obtained after Higgs, electroweak, flavor and low-energy selections. The numerical results identify $h\to sb$, $t\to ch$, $h\to db$, and $h\to\mu\tau$ as the leading channels in the numerical setup used here, reveal a structured flavor pattern, and show that the down-sector Higgs FCNC rates are much more stable at tree level than the texture- and alignment-sensitive tree-level $t\to ch$ and $h\to\mu\tau$ modes. We also examine one-loop vertex effects in Higgs-mediated FCNC decays and rare top decays.

The paper is organized as follows. Section~\ref{sec:model-setup} describes the Type 1B F2HDM. Section~\ref{sec:observables-method} defines the observables and numerical procedure. Section~\ref{sec:numerical-analysis} presents the numerical results. Section~\ref{sec:conclusions} gives the conclusions.

\section{The Type 1B F2HDM}
\label{sec:model-setup}

We work in the Type 1B realization of the flavorful two-Higgs-doublet model with a twist~\cite{Altmannshofer:2018bch}. This model is different from the ordinary natural-flavor-conserving Type-I two-Higgs-doublet model. In the Type 1B F2HDM, the two scalar doublets have a non-standard flavor assignment: \(\Phi_1\) gives the dominant contribution to fermion masses of the first two generations, whereas \(\Phi_2\) gives fermion masses dominantly of the third generation. Consequently, the two doublets couple to different fermion species dominantly, and tree-level flavor-changing neutral Higgs couplings are generically present.

The Type 1B assignment used here is
\begin{equation}
\begin{array}{c|ccc}
  & u & d & \ell \\
\hline
  \hbox{first two generations} & \Phi_1 & \Phi_1 & \Phi_1 \\
  \hbox{third generation}      & \Phi_2 & \Phi_2 & \Phi_2
\end{array}
\label{eq:type1b-assignment}
\end{equation}
understood as a texture assignment distinct from a natural-flavor-conserving zero-pattern.  The small entries controlled by the approximate \(U(2)^5\) flavor structure and by the \(O(1)\) coefficients populate both doublets in a way that reproduces fermion masses and the CKM matrix.

\subsection{Scalar sector}
\label{subsec:scalar-sector}

The scalar sector contains two scalar doublets \(\Phi_a\sim(2,+1/2)\) under \(SU(2)_L\times U(1)_Y\),
\begin{equation}
  \Phi_a =
  \begin{pmatrix}
    \phi_a^+ \\
    \dfrac{v_a+\rho_a+i\eta_a}{\sqrt{2}}
  \end{pmatrix},
  \qquad a=1,2,
\end{equation}
with
\begin{equation}
  v^2=v_1^2+v_2^2,\qquad \tan\beta=\frac{v_2}{v_1}.
\end{equation}
We restrict the scalar sector to the commonly used CP-conserving softly broken
\(Z_2\)-type scalar-potential form. The \(Z_2\)-breaking is retained only
through the quadratic soft term \(m_{12}^2\), while the hard-breaking quartic
terms are not included. The scalar potential is
\begin{align}
V(\Phi_1,\Phi_2)={}&
  m_{11}^2\,\Phi_1^\dagger\Phi_1
 +m_{22}^2\,\Phi_2^\dagger\Phi_2
 -\left[m_{12}^2\,\Phi_1^\dagger\Phi_2+\mathrm{h.c.}\right]
\nonumber\\
&+\lambda_1(\Phi_1^\dagger\Phi_1)^2
 +\lambda_2(\Phi_2^\dagger\Phi_2)^2
 +\lambda_3(\Phi_1^\dagger\Phi_1)(\Phi_2^\dagger\Phi_2)
 +\lambda_4(\Phi_2^\dagger\Phi_1)(\Phi_1^\dagger\Phi_2)
\nonumber\\
&+\left[
 \frac{\lambda_5}{2}(\Phi_2^\dagger\Phi_1)^2
 +\mathrm{h.c.}\right] .
\label{eq:scalar-potential}
\end{align}
All scalar-potential parameters in Eq.~\eqref{eq:scalar-potential} are taken real in the CP-conserving numerical implementation.  The neutral CP-even states are obtained from
\begin{equation}
  \begin{pmatrix} h \\ H \end{pmatrix}
  = Z^h
  \begin{pmatrix} \rho_1 \\ \rho_2 \end{pmatrix},
\end{equation}
with analogous rotations for the CP-odd and charged sectors. The reduced light-Higgs vector coupling is
\begin{equation}
\kappa_V\equiv
\frac{g_{hVV}}{g^{\rm SM}_{hVV}}
=
\frac{v_1 Z^h_{11}+v_2 Z^h_{12}}{v}.
\label{eq:kappa-v-definition}
\end{equation}
The alignment preselection used below is expressed in terms of this quantity.  The alignment-distance variable used in the plots is
\begin{equation*}
|\cos(\beta-\alpha)|\equiv \sqrt{\max\!
\left(0,1-\kappa_V^2\right)} .
\end{equation*}
which follows the convention \(\kappa_V=\sin(\beta-\alpha)\) for a SM-like light Higgs. The sign of the soft term in Eq.~\eqref{eq:scalar-potential} follows the project convention \(-m_{12}^2(\Phi_1^\dagger\Phi_2+\mathrm{h.c.})\).

\subsection{Yukawa Lagrangian and Type 1B texture}
\label{subsec:yukawa-lagrangian}

The two-doublet notation used in this work is \(\Phi_{1,2}\).  In the mass-parameter notation of Ref.~\cite{Altmannshofer:2018bch}, the doublet associated with the dominant third-generation masses is denoted by \(\phi\), while the other doublet is denoted by \(\phi^\prime\).  For the Type 1B assignment in Eq.~\eqref{eq:type1b-assignment}, these conventions are related by
\[
\phi\equiv\Phi_2,
\qquad
\phi^\prime\equiv\Phi_1,
\qquad
v=v_2,
\qquad
v'=v_1,
\]
and the corresponding Yukawa matrices are related by
\[
\lambda^u=Y_2^u,
\quad
\lambda^d=Y_2^d,
\quad
\lambda^e=Y_2^\ell,
\quad
\lambda^{\prime u}=Y_1^u,
\quad
\lambda^{\prime d}=Y_1^d,
\quad
\lambda^{\prime e}=Y_1^\ell .
\]
With \(\widetilde\phi^{(\prime)}=i\sigma_2(\phi^{(\prime)})^*\), the Yukawa part of the flavorful two-Higgs-doublet Lagrangian is written as
\begin{align}
-\mathcal L_Y\supset{}&
\sum_{i,j}\left[
\lambda^u_{ij}(\bar q_i u_j)\widetilde\phi
+\lambda^d_{ij}(\bar q_i d_j)\phi
+\lambda^e_{ij}(\bar \ell_i e_j)\phi
+\mathrm{h.c.}\right]
\nonumber\\
&+\sum_{i,j}\left[
\lambda^{\prime u}_{ij}(\bar q_i u_j)\widetilde\phi^{\,\prime}
+\lambda^{\prime d}_{ij}(\bar q_i d_j)\phi^\prime
+\lambda^{\prime e}_{ij}(\bar \ell_i e_j)\phi^\prime
+\mathrm{h.c.}\right] .
\label{eq:yukawa-lagrangian}
\end{align}
Together with the mapping above, Eq.~\eqref{eq:yukawa-lagrangian} serves as the notation bridge between the \(\phi,\phi^\prime\) mass-parameter convention and the \(\Phi_1,\Phi_2\) texture notation used below.

The Type 1B flavor assignment is represented by the following flavor-basis Yukawa textures.  Up to independent \(O(1)\) coefficients, the first doublet carries the light-generation texture, while the second doublet carries the rank-one third-generation texture~\cite{Altmannshofer:2018bch}:
\begin{align}
Y^u_1 &\sim \frac{\sqrt2}{v_1}
\begin{pmatrix}
 m_u & m_u & m_u \\
 m_u & m_c & m_c \\
 m_u & m_c & m_c
\end{pmatrix},
&
Y^u_2 &\sim \frac{\sqrt2}{v_2}
\begin{pmatrix}
0&0&0\\
0&0&0\\
0&0&m_t
\end{pmatrix}
\label{eq:type1b-texture-up}
\\[1ex]
Y^d_1 &\sim \frac{\sqrt2}{v_1}
\begin{pmatrix}
 m_d & \lambda m_s & \lambda^3 m_b \\
 m_d & m_s & \lambda^2 m_b \\
 m_d & m_s & m_s
\end{pmatrix},
&
Y^d_2 &\sim \frac{\sqrt2}{v_2}
\begin{pmatrix}
0&0&0\\
0&0&0\\
0&0&m_b
\end{pmatrix}
\label{eq:type1b-texture-down}
\\[1ex]
Y^\ell_1 &\sim \frac{\sqrt2}{v_1}
\begin{pmatrix}
 m_e & m_e & m_e \\
 m_e & m_\mu & m_\mu \\
 m_e & m_\mu & m_\mu
\end{pmatrix},
&
Y^\ell_2 &\sim \frac{\sqrt2}{v_2}
\begin{pmatrix}
0&0&0\\
0&0&0\\
0&0&m_\tau
\end{pmatrix}
\label{eq:type1b-texture-lepton}
\end{align}
Here \(\lambda\) denotes the Cabibbo-sized expansion parameter in the down sector, and the symbol \(\sim\) indicates the parametric texture form.  The explicit scan uses independent coefficients \(O^u_{ij}\), \(O^d_{ij}\), and \(O^\ell_{ij}\) to realize these textures while exactly reconstructing the physical fermion masses.

Using this mapping and rotating the fermions into mass eigenstates, we define
\begin{equation}
 m^u_{q q'}=
 \frac{v}{\sqrt2}\,
 \left\langle q_L\middle|\lambda^u\middle|q'_R\right\rangle,
 \qquad
 m^{\prime u}_{q q'}=
 \frac{v'}{\sqrt2}\,
 \left\langle q_L\middle|\lambda^{\prime u}\middle|q'_R\right\rangle,
 \qquad q,q'=u,c,t ,
\label{eq:mass-param-up}
\end{equation}
\begin{equation}
 m^d_{q q'}=
 \frac{v}{\sqrt2}\,
 \left\langle q_L\middle|\lambda^d\middle|q'_R\right\rangle,
 \qquad
 m^{\prime d}_{q q'}=
 \frac{v'}{\sqrt2}\,
 \left\langle q_L\middle|\lambda^{\prime d}\middle|q'_R\right\rangle,
 \qquad q,q'=d,s,b ,
\label{eq:mass-param-down}
\end{equation}
\begin{equation}
 m^\ell_{\ell \ell'}=
 \frac{v}{\sqrt2}\,
 \left\langle \ell_L\middle|\lambda^\ell\middle|\ell'_R\right\rangle,
 \qquad
 m^{\prime \ell}_{\ell \ell'}=
 \frac{v'}{\sqrt2}\,
 \left\langle \ell_L\middle|\lambda^{\prime \ell}\middle|\ell'_R\right\rangle,
 \qquad \ell,\ell'=e,\mu,\tau .
\label{eq:mass-param-lepton}
\end{equation}
In each charged-fermion sector these mass parameters obey
\(m^f_{ij}+m^{\prime f}_{ij}=m_{f_i}\delta_{ij}\).  The charged-fermion
texture coefficients are denoted as \(O^u_{ij}\), \(O^d_{ij}\), and
\(O^\ell_{ij}\); the flavor label is a superscript and the matrix element is a
subscript.

The explicit primed mass matrices \(M'_f\equiv(m^{\prime f}_{ij})\) used to reconstruct the charged-fermion Yukawa sector are~\cite{Altmannshofer:2018bch}
\begin{equation}
M_u'=
\begin{pmatrix}
 m_u+O^u_{11}\dfrac{m_u^2}{m_t} &
 \dfrac{(O^u_{13}m_u)(O^u_{32}m_c)}{m_t}\left(1+O^u_{12}\dfrac{m_c}{m_t}\right) &
 O^u_{13}m_u \\
 \dfrac{(O^u_{23}m_c)(O^u_{31}m_u)}{m_t}\left(1+O^u_{21}\dfrac{m_c}{m_t}\right) &
 m_c+O^u_{22}\dfrac{m_c^2}{m_t} &
 O^u_{23}m_c \\
 O^u_{31}m_u &
 O^u_{32}m_c &
 O^u_{33}m_c
\end{pmatrix} .
\label{eq:mprime-up}
\end{equation}
\begin{equation}
\resizebox{0.98\textwidth}{!}{\(\displaystyle
M_d'=\begin{pmatrix}
 m_d-O^d_{31}m_d V_{td}^*\left(1+O^d_{11}\dfrac{m_s}{m_b}\right) &
 O^d_{32}m_s V_{td}^*\left(1+O^d_{12}\dfrac{m_s}{m_b}\right) &
 -V_{td}^*m_b\left(1+O^d_{13}\dfrac{m_s}{m_b}\right) \\
 O^d_{31}m_d V_{ts}^*\left(1+O^d_{21}\dfrac{m_s}{m_b}\right) &
 m_s-O^d_{32}m_s V_{ts}^*\left(1+O^d_{22}\dfrac{m_s}{m_b}\right) &
 -V_{ts}^*m_b\left(1+O^d_{23}\dfrac{m_s}{m_b}\right) \\
 O^d_{31}m_d &
 O^d_{32}m_s &
 O^d_{33}m_s
\end{pmatrix} .
\)}
\label{eq:mprime-down}
\end{equation}
\begin{equation}
M_\ell'=\begin{pmatrix}
 m_e+O^\ell_{11}\dfrac{m_e^2}{m_\tau} &
 \dfrac{(O^\ell_{13}m_e)(O^\ell_{32}m_\mu)}{m_\tau}\left(1+O^\ell_{12}\dfrac{m_\mu}{m_\tau}\right) &
 O^\ell_{13}m_e \\
 \dfrac{(O^\ell_{23}m_\mu)(O^\ell_{31}m_e)}{m_\tau}\left(1+O^\ell_{21}\dfrac{m_\mu}{m_\tau}\right) &
 m_\mu+O^\ell_{22}\dfrac{m_\mu^2}{m_\tau} &
 O^\ell_{23}m_\mu \\
 O^\ell_{31}m_e &
 O^\ell_{32}m_\mu &
 O^\ell_{33}m_\mu
\end{pmatrix} .
\label{eq:mprime-lepton}
\end{equation}
Together with Eqs.~\eqref{eq:mass-param-up}--\eqref{eq:mass-param-lepton}, these matrices determine the Yukawa couplings used in the charged-fermion FCNC analysis.

In the numerical realization used below, the CKM matrix is generated in the down sector,
\begin{equation}
  V_{Lu}\simeq I,
  \qquad
  V_{Ru}\simeq I,
  \qquad
  V_{Ld}\simeq V_{\rm CKM},
  \qquad
  V_{Rd}\simeq I .
\label{eq:ckm-down-convention}
\end{equation}
Equation~\eqref{eq:ckm-down-convention} is the basis choice and numerical ansatz used in the scan.  The texture by itself does not fix this rotation pattern.  With our fermion-rotation convention,
\begin{equation*}
V_{\rm CKM}=V_{Lu}^{\dagger}V_{Ld} .
\end{equation*}
Thus \(V_{Lu}\simeq I\) implies \(V_{Ld}\simeq V_{\rm CKM}\), while \(V_{Rd}\simeq I\) is an additional simplifying assumption.
In this section we specify the Type 1B texture and the corresponding mass-basis Yukawa definition. The conversion of these mass-basis matrices into the numerical input basis used for the scan is part of the numerical procedure.

\section{Observables and numerical procedure}
\label{sec:observables-method}

\subsection{Decay channels and charge-conjugate convention}
\label{subsec:channels-source}

The numerical analysis considers the flavor-violating Higgs channels
\(h\to\mu\tau\), \(h\to e\tau\), \(h\to e\mu\),
\(h\to sb\), \(h\to db\), and \(h\to uc\), together with the top FCNC channels
\(t\to ch\) and \(t\to uh\).

Throughout the paper, the shorthand notation for neutral Higgs decays includes the charge-conjugate final states. We define
\begin{align}
\BR(h\to sb)&\equiv \BR(h\to s\bar b)+\BR(h\to \bar s b),\\
\BR(h\to db)&\equiv \BR(h\to d\bar b)+\BR(h\to \bar d b),\\
\BR(h\to uc)&\equiv \BR(h\to u\bar c)+\BR(h\to \bar{u}c),
\end{align}
and, for charged-lepton channels,
\begin{align}
\BR(h\to \mu\tau)&\equiv \BR(h\to \mu^-\tau^+)+\BR(h\to \mu^+\tau^-),\\
\BR(h\to e\tau)&\equiv \BR(h\to e^-\tau^+)+\BR(h\to e^+\tau^-),\\
\BR(h\to e\mu)&\equiv \BR(h\to e^-\mu^+)+\BR(h\to e^+\mu^-).
\end{align}
The same charge-conjugate convention is used when comparing tree-level and one-loop quantities for the quark channels.

\subsection{Amplitudes and decay widths}
\label{subsec:amplitudes-widths}

For the quark-Higgs channels, we parameterize the amplitude in a chiral form-factor basis,
\begin{equation}
\mathcal M(h\to f_i\bar f_j)
=
\bar u_i(p_i)
\left[
G^{h,ij}_{L}P_L+
G^{h,ij}_{R}P_R
\right]
v_j(p_j),
\label{eq:hff-amplitude}
\end{equation}
where \(f_i\bar f_j=s\bar b,d\bar b,u\bar c\) for the explicitly calculated directions. The one-loop form factors are decomposed schematically as
\begin{equation}
G^{h,ij}_{X}
=
g^{h,ij}_{X,\mathrm{tree}}
+
\Delta g^{h,ij}_{X,\mathrm{tri}},
\qquad X=L,R,
\label{eq:hff-formfactor-decomp}
\end{equation}
where \(\Delta g_{X,\mathrm{tri}}\) denotes the one-loop vertex contribution.

For top FCNC decays, we use
\begin{equation}
\mathcal M(t\to q_i h)
=
\bar u_{q_i}(p_q)
\left[
G^{tq_i}_{L}P_L+
G^{tq_i}_{R}P_R
\right]
u_t(p_t),
\label{eq:tqh-amplitude}
\end{equation}
with
\begin{equation}
G^{tq_i}_{X}
=
g^{tq_i}_{X,\mathrm{tree}}
+
\Delta g^{tq_i}_{X,\mathrm{tri}},
\qquad q_i=u,c,
\qquad X=L,R.
\label{eq:tqh-formfactor-decomp}
\end{equation}

With this convention, the two-body decay width for a neutral scalar decay into two fermions is
\begin{equation}
\Gamma(h\to f_i\bar f_j)
=
\frac{N_c\,\lambda^{1/2}(m_h^2,m_i^2,m_j^2)}
{16\pi m_h^3}
\left[
(m_h^2-m_i^2-m_j^2)(|G_L|^2+|G_R|^2)
-4m_i m_j\,{\rm Re}(G_LG_R^\ast)
\right],
\label{eq:hff-width}
\end{equation}
where \(N_c=3\) for quark final states and
\(\lambda(x,y,z)=x^2+y^2+z^2-2xy-2xz-2yz\). The charge-conjugate branching ratio is
\begin{equation}
\BR(h\to f_i f_j)
=
\frac{
\Gamma(h\to f_i\bar f_j)+\Gamma(h\to \bar f_i f_j)
}
{\Gamma_h^{\rm tot}}.
\label{eq:hff-br}
\end{equation}
For the top decay in Eq.~\eqref{eq:tqh-amplitude},
\begin{equation}
\Gamma(t\to q_i h)
=
\frac{\lambda^{1/2}(m_t^2,m_q^2,m_h^2)}
{16\pi m_t^3}
\left[
(m_t^2+m_q^2-m_h^2)(|G_L|^2+|G_R|^2)
+4m_t m_q\,{\rm Re}(G_LG_R^\ast)
\right].
\label{eq:tqh-width}
\end{equation}
The branching ratio is
\begin{equation}
\BR(t\to q_i h)=\frac{\Gamma(t\to q_i h)}{\Gamma_t^{\rm tot}}.
\label{eq:tqh-br}
\end{equation}
The total widths \(\Gamma_h^{\rm tot}\) and \(\Gamma_t^{\rm tot}\) are the widths used in the corresponding numerical normalization.

\subsection{One-loop vertex corrections}
\label{subsec:one-loop-vertex}

\begin{figure}[!htbp]
  \centering
  \begin{subfigure}[t]{0.72\textwidth}
    \centering
    \includegraphics[width=\textwidth]{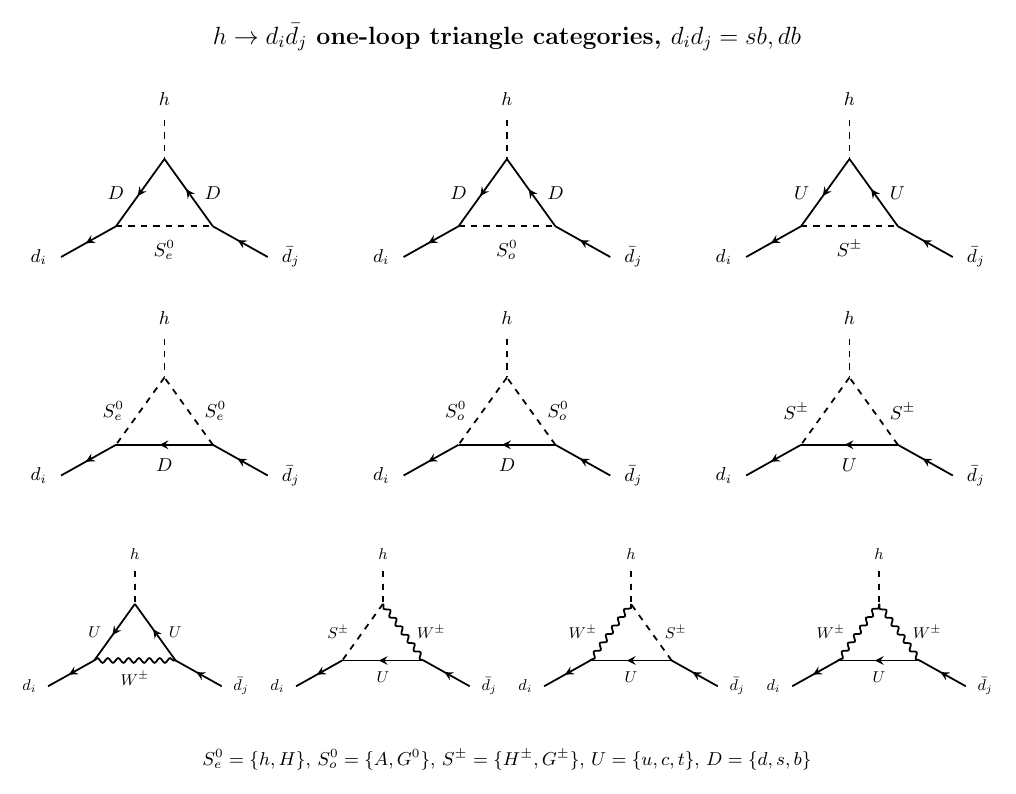}
    \caption{Down-sector Higgs FCNC channels $h\to d_i\bar d_j$ with $d_id_j=sb,db$.}
    \label{fig:one-loop-triangle-hdd}
  \end{subfigure}

  \vspace{0.3em}
  \begin{subfigure}[t]{0.72\textwidth}
    \centering
    \includegraphics[width=\textwidth]{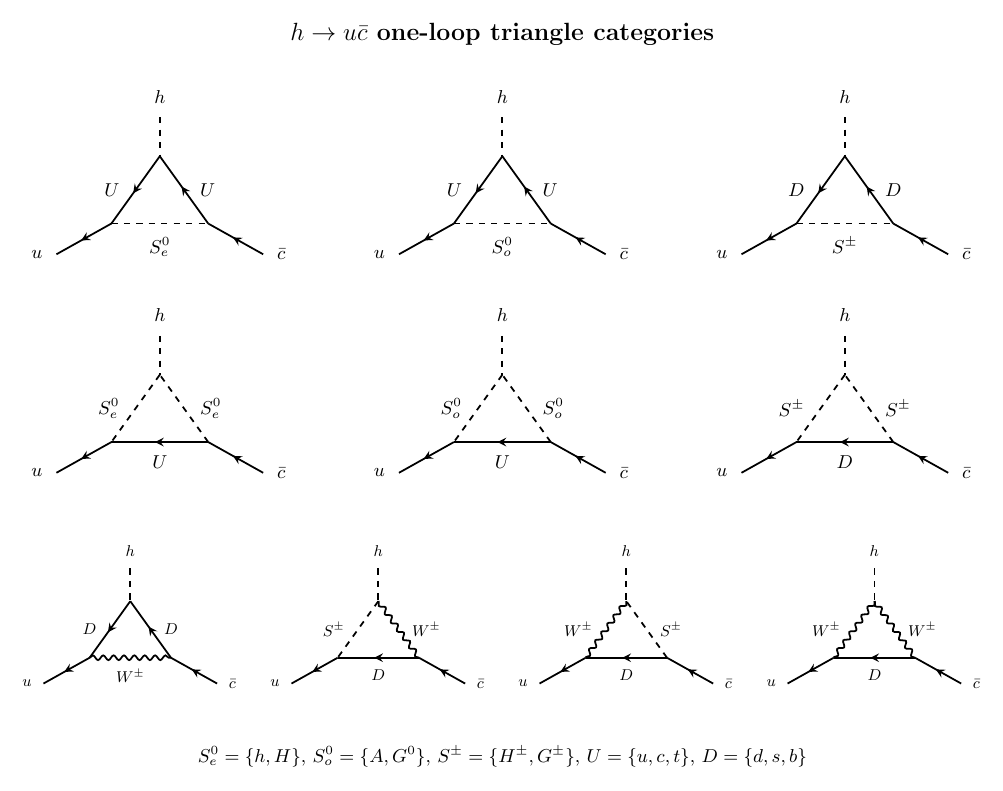}
    \caption{Up-sector Higgs FCNC channel $h\to u\bar c$.}
    \label{fig:one-loop-triangle-huc}
  \end{subfigure}
  \caption{One-loop triangle categories for the Higgs FCNC channels.  The notation is $S_e^0=\{h,H\}$, $S_o^0=\{A,G^0\}$, $S^\pm=\{H^\pm,G^\pm\}$, with $U=\{u,c,t\}$ and $D=\{d,s,b\}$.}
\end{figure}

\begin{figure}[!htbp]
  \centering
  \includegraphics[width=0.76\textwidth]{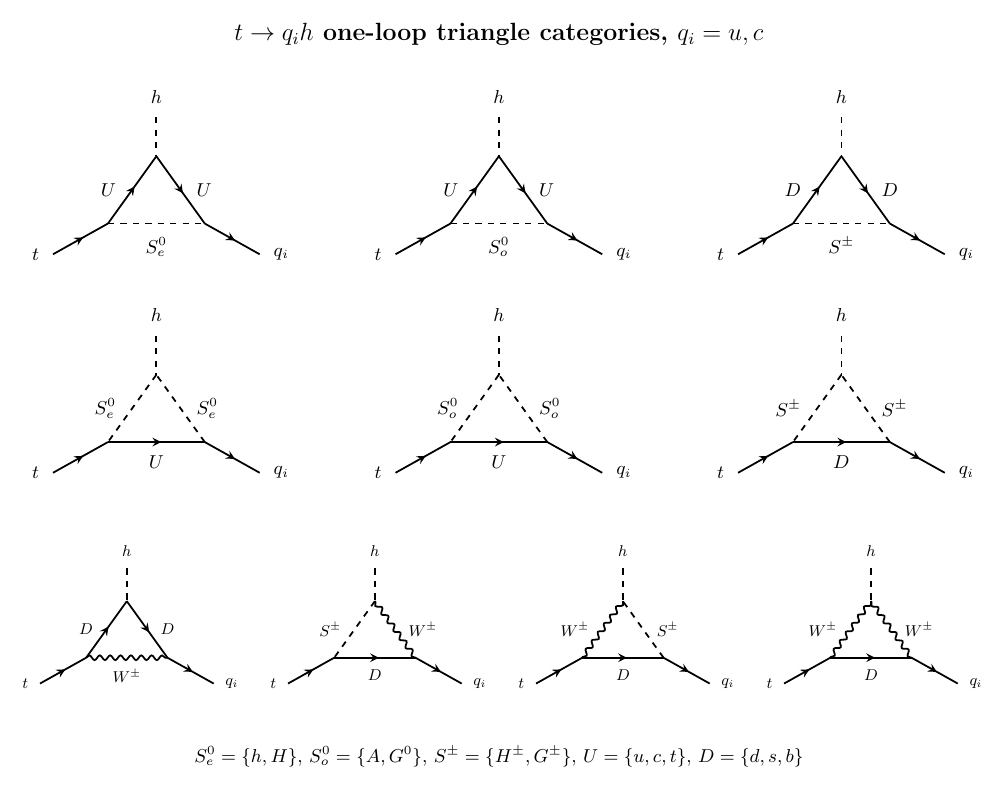}
  \caption{One-loop triangle categories for the top FCNC channels $t\to q_i h$ with $q_i=u,c$.  The notation is the same as in Fig.~\ref{fig:one-loop-triangle-hdd}.}
  \label{fig:one-loop-triangle-tqh}
\end{figure}

For the explicitly calculated quark-Higgs and top-Higgs FCNC channels, we compare
the tree-level amplitude with the one-loop vertex contribution shown by category
in Figs.~\ref{fig:one-loop-triangle-hdd}--\ref{fig:one-loop-triangle-tqh}.  We write the amplitude as
\begin{equation}
\mathcal M_{\mathrm{1L}}
=
\mathcal M_{\rm tree}
+
\mathcal M_{\rm loop} .
\label{eq:one-loop-amplitude}
\end{equation}
The corresponding rate-level expression is
\begin{equation}
\BR_{\mathrm{1L}}
=
\BR_{\rm tree}
+
\Delta\BR_{\rm int}
+
\BR_{\rm loop^2},
\label{eq:one-loop-br-decomp}
\end{equation}
where \(\Delta\BR_{\rm int}\) is proportional to
\(2\mathrm{Re}(\mathcal M_{\rm tree}^{*}\mathcal M_{\rm loop})\), and
\(\BR_{\rm loop^2}\) is proportional to \(|\mathcal M_{\rm loop}|^2\).  We define
\begin{equation}
\delta_{\mathrm{1L}}
\equiv
\frac{\BR_{\mathrm{1L}}-\BR_{\rm tree}}{\BR_{\rm tree}} .
\label{eq:one-loop-delta}
\end{equation}

\subsection{Parameter scan and constraints}
\label{subsec:scan-constraints}

The numerical scan uses the parameter ranges and preselection requirements summarized in Table~\ref{tab:scan-ranges}. The scalar sector is sampled in \(\tan\beta\), \(m_{12}^2\), and the quartic couplings \(\lambda_{1,\ldots,5}\) appearing in Eq.~\eqref{eq:scalar-potential}. This softly broken \(Z_2\)-type scalar-potential choice removes hard \(Z_2\)-breaking quartic terms, reduces the number of independent scalar-sector parameters, and keeps the flavor non-universality studied here in the Yukawa texture rather than in additional scalar-potential structures. The scan further imposes a SM-like light Higgs mass window, heavy non-SM scalar masses above the TeV scale, a charged--pseudoscalar mass-splitting requirement, and an alignment preselection before the full set of phenomenological constraints is applied.

\begin{table}[!htbp]
\centering
\small
\begin{tabular}{|l|l|}
\hline
Parameter or requirement & Range or selection \\
\hline
$\tan\beta=v_2/v_1$ & $1\le \tan\beta\le 30$ \\
$m_{12}^2$ & $(2.5\times10^5,\,1.0\times10^6)~{\rm GeV}^2$ \\
$\lambda_1$ & $0\le \lambda_1\le 3$ \\
$\lambda_2$ & $0.06\le \lambda_2\le 0.13$ \\
$\lambda_{3,4,5}$ & $-5\le \lambda_{3,4,5}\le 5$ \\
Light Higgs mass & $124~{\rm GeV}<m_h<126~{\rm GeV}$ \\
Heavy scalar masses & $m_H,m_A,m_{H^\pm}>1~{\rm TeV}$ \\
Charged--pseudoscalar splitting & $|m_A-m_{H^\pm}|<200~{\rm GeV}$ \\
Alignment proxy & $|\kappa_V|^2>0.95$ \\
$O^u_{ij}$ & $0.1\le |O^u_{ij}|\le 10$ \\
$O^d_{ii}$ & $0.01\le |O^d_{ii}|\le 10$ \\
$O^d_{12,21}$, $O^d_{13,31}$ & $0.01\le |O^d_{ij}|\le 6$ \\
$O^d_{23,32}$ & $0.01\le |O^d_{23,32}|\le 1$ \\
$O^\ell_{ij}$ & $1\le |O^\ell_{ij}|\le 10$ \\
Texture-factor phases & $0$ \\
\hline
\end{tabular}
\caption{Input parameter ranges and preselection requirements for the Type 1B F2HDM scan.}
\label{tab:scan-ranges}

\end{table}

We impose the following theoretical and phenomenological constraints in the numerical pipeline. Theoretical constraints on the scalar sector are imposed following the standard 2HDM perturbativity, tree-level unitarity and bounded-from-below criteria, translated to the scalar-potential convention of Eq.~\eqref{eq:scalar-potential} \cite{Branco:2011iw}.

\paragraph{Perturbativity.}
The scalar quartic couplings are required to satisfy
\(\max |\lambda_i^{\rm ref}|<4\pi\).  The reconstructed Yukawa matrices are also required to remain perturbative,
\(\max |Y^f_{a,ij}|\le \sqrt{4\pi}\).

\paragraph{Tree-level scalar unitarity.}
The largest eigenvalue entering the scalar two-body scattering matrix is required to be smaller than \(8\pi\).

\paragraph{Bounded-from-below and vacuum stability.}
Our scalar potential uses
\(\lambda_1(\Phi_1^\dagger\Phi_1)^2+\lambda_2(\Phi_2^\dagger\Phi_2)^2\),
while the convention of Ref.~\cite{Branco:2011iw} uses
\((\lambda_1^B/2)(\Phi_1^\dagger\Phi_1)^2+(\lambda_2^B/2)(\Phi_2^\dagger\Phi_2)^2\).
The correspondence is
\(\lambda_1^B=2\lambda_1\), \(\lambda_2^B=2\lambda_2\), while
\(\lambda_{3,4,5}^B=\lambda_{3,4,5}\) up to the harmless
\(\Phi_1^\dagger\Phi_2\) versus \(\Phi_2^\dagger\Phi_1\) complex-conjugation convention.
For the CP-conserving scalar potential in Eq.~\eqref{eq:scalar-potential}, the bounded-from-below conditions become
\begin{align}
\lambda_1&>0, &
\lambda_2&>0,\nonumber\\
\lambda_3+2\sqrt{\lambda_1\lambda_2}&>0, &
\lambda_3+\lambda_4-|\lambda_5|+2\sqrt{\lambda_1\lambda_2}&>0 .
\end{align}

\paragraph{Higgs signal strengths.}
The Higgs signal-strength selection uses the global one-sigma condition
\begin{equation}
\Delta\chi^2_{\rm HS}
\equiv \chi^2_{\rm HS}-\chi^2_{{\rm HS},\min}\leq 1,
\label{eq:higgstools-rule}
\end{equation}
where \(\chi^2_{\rm HS}\) is the HiggsSignals chi-square and \(\chi^2_{{\rm HS},\min}\) is the minimum value in the generated scan.  Numerically, the 50000-point sample gives \(\chi^2_{{\rm HS},\min}=149.872\), so the applied cut is \(\chi^2_{\rm HS}\le150.872\).

\paragraph{Electroweak precision constraints.}
We require the obtained \(S,T,U\) values to lie within the current experimental \(1\sigma\) ranges quoted in the Review of Particle Physics~\cite{ParticleDataGroup:2024cfk}:
\begin{equation}
S=-0.04\pm0.10,\qquad
T=0.01\pm0.12,\qquad
U=-0.01\pm0.09 .
\end{equation}
The three intervals are applied independently in the numerical scan.

\paragraph{Neutral-meson mixing.}
Neutral-meson mixing constraints are imposed using the observable-specific
two-sided 95\% probability intervals reported in Table~3 of the
model-independent analysis of Ref.~\cite{UTfit:2007eik}:
\begin{align}
0.63 \leq \frac{\Delta M_{B_s}}{\Delta M_{B_s}^{\rm SM}}&\leq 2.07, &
0.53 \leq \frac{\Delta M_{B_d}}{\Delta M_{B_d}^{\rm SM}}&\leq 2.05,\nonumber\\
0.51 \leq \frac{\Delta M_K}{\Delta M_K^{\rm SM}}&\leq 2.07, &
0.66 \leq \frac{\epsilon_K}{\epsilon_K^{\rm SM}}&\leq 1.31 .
\label{eq:meson-mixing}
\end{align}
The four conditions are applied independently. The $B_s$ and $B_d$ ratios
constrain neutral-meson mixing in the $b\leftrightarrow s$ and
$b\leftrightarrow d$ sectors, respectively, while $\Delta M_K$ and
$\epsilon_K$ probe the dispersive and CP-violating components of
$K^0$--$\bar K^0$ mixing. For $\Delta M_K$, we use the short-distance ratio
returned by FlavorKit rather than imposing the experimental mass-splitting
uncertainty directly, owing to the sizable long-distance contribution to
kaon mixing.

\paragraph{\(B\to X_s\gamma\).}
The radiative \(B\)-decay constraint is imposed using the inclusive reference interval~\cite{HFLAV:2016hnz} as
\begin{equation}
2.99\times10^{-4}\le \BR(B\to X_s\gamma)\le 3.87\times10^{-4}.
\end{equation}

\paragraph{Charged-lepton flavor and \(\mu-e\) conversion constraints.}
The lepton-sector HFV branching ratios are required to satisfy the current direct \(95\%\) C.L. limits
\begin{align}
\BR(h\to e\mu)&<4.4\times10^{-5}, &
\BR(h\to e\tau)&<2.0\times10^{-3}, &
\BR(h\to\mu\tau)&<1.5\times10^{-3}.
\end{align}
These entries correspond to the observed limits for \(h\to e\mu\)~\cite{CMS:2023xpx}, \(h\to e\tau\)~\cite{ATLAS:2023mvd}, and \(h\to\mu\tau\)~\cite{CMS:2021rsq}, respectively.
The \(\mu-e\) conversion constraints are imposed as~\cite{Dohmen:1993mp,Bertl:2006up}
\begin{align}
{\rm CR}_{\rm Ti}&<4.3\times10^{-12}, &
{\rm CR}_{\rm Au}&<7.0\times10^{-13}.
\end{align}

\paragraph{Additional CLFV observables.}
Radiative and three-body CLFV observables are evaluated as auxiliary predictions outside the independent sample-reducing cut-flow. In the surviving sample, these observables remain below their corresponding current experimental limits. The largest ratios of the predicted values to the corresponding experimental upper bounds are approximately \(3.2\times10^{-2}\) for \(\mu\to e\gamma\) and \(5.2\times10^{-4}\) for \(\tau\to3\mu\), with smaller ratios in the other radiative and three-body CLFV channels. Top-FCNC and Higgs-FCNC quantities are evaluated for interpretation outside the selection cuts.

After all imposed selections, the final sample is fixed for the numerical analysis below.

\subsection{Computational workflow}
\label{subsec:software-workflow}

The model implementation and SPheno interface are generated with SARAH 4.15.4~\cite{Staub:2013tta}. The particle spectrum, low-energy observables, and lepton-sector HFV Higgs rates are evaluated with SPheno 4.0.6~\cite{Porod:2003um,Porod:2011nf}. In particular, the neutral-meson-mixing observables are computed using the SARAH-generated SPheno/FlavorKit routines~\cite{Porod:2014xia}. Higgs signal-strength constraints are evaluated with the HiggsSignals functionality used in the HiggsTools 1.1.3 run~\cite{Bechtle:2013xfa,Bechtle:2020uwn}. The FCNC one-loop vertex corrections are obtained using FeynArts 3.12, FormCalc 9.10 and LoopTools 2.16~\cite{Hahn:2000kx,Hahn:1998yk,Hahn:2000jm}, with SARAH-matched $G_{H^+ud}^{L,R}$ conjugations.

The numerical chain proceeds as follows. Scalar-sector points satisfying the theoretical and Higgs-sector preselection requirements are first generated. The charged-fermion Yukawa textures are then reconstructed in the mass and gauge bases according to the Type 1B flavor structure, with the CKM rotation assigned to the down sector. The Higgs signal-strength selection is applied next, followed by the precision, flavor, and flavor-violating observable requirements. The final branching-ratio distributions and hierarchy observables are then assembled using the conventions of Sec.~\ref{subsec:channels-source}.

\section{Numerical analysis}
\label{sec:numerical-analysis}

The scan ranges and observable definitions were described in Sec.~\ref{sec:observables-method}. In this section we focus on the numerical consequences for the surviving Type 1B F2HDM sample.

\subsection{Cut-flow and final sample}
\label{subsec:cutflow}

Starting from 50000 generated points in the parameter region summarized in Table~\ref{tab:scan-ranges}, 6790 points pass the global HiggsSignals condition \(\Delta\chi^2_{\rm HS}\le1\).  Within this subset, the meson-mixing constraints reduce the sample to 1094 points, the $B\to X_s\gamma$ selection to 552 points, and the electroweak precision selection to 542 points. The direct HFV and $\mu$--$e$ conversion constraints do not further reduce this sample, leaving
\begin{equation}
N_{\rm final}=542 .
\end{equation}
These points form the constrained surviving sample used in the following analysis.

\begin{figure}[!htbp]
  \centering
  \includegraphics[width=0.70\textwidth]{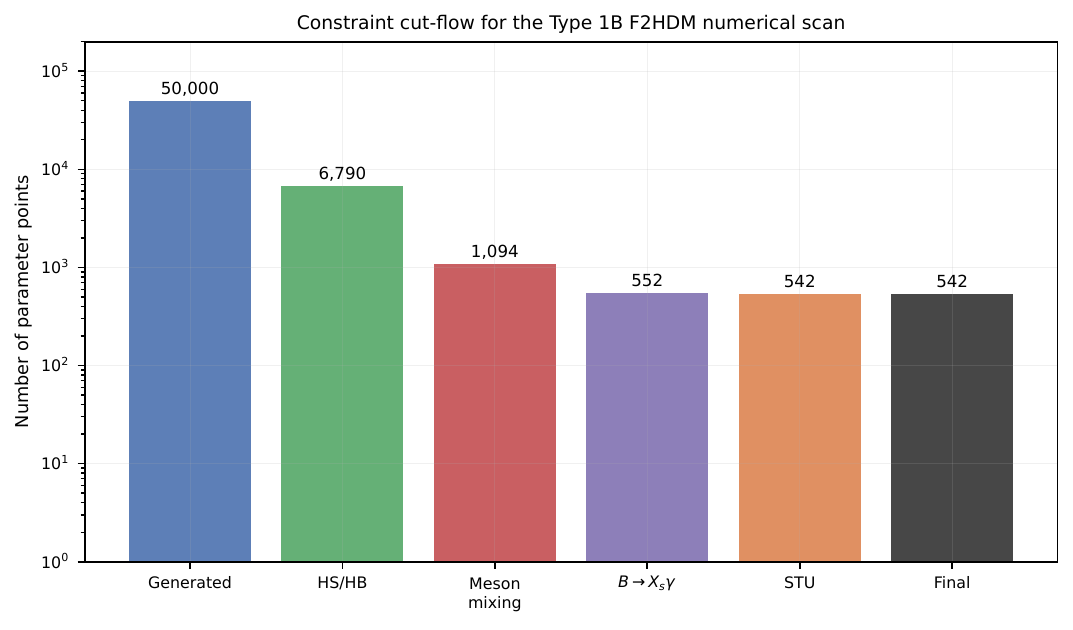}
  \caption{Cut-flow for the Type 1B F2HDM numerical scan after the Higgs, flavor, and electroweak selections.}
  \label{fig:cutflow}
\end{figure}

\FloatBarrier

\paragraph*{One-loop rate update and branching-ratio convention.}

Before discussing the hierarchy and parameter dependences, we specify the rate convention used in the numerical plots.  The surviving sample is defined by the cut-flow above; the one-loop vertex corrections are then evaluated on this fixed sample and are not used as additional selection cuts.  For the Higgs-mediated FCNC decays and rare top decays we use the rate definition of Sec.~\ref{subsec:one-loop-vertex}, namely the tree-level contribution plus the tree--loop interference and the squared one-loop amplitude.  Thus, unless explicitly stated otherwise, the quoted rates for $h\to sb$, $h\to db$, $h\to uc$, $t\to ch$, and $t\to uh$ below refer to the one-loop-updated branching ratios $\BR_{\mathrm{1L}}$.

For the lepton-sector HFV Higgs channels, the numerical rates are those obtained from the SPheno loop calculation in the workflow of Sec.~\ref{subsec:software-workflow}.  Tree-level quantities are shown separately only when the text explicitly discusses tree-level stability.  With this convention, the hierarchy plots in Figs.~\ref{fig:br-overview}--\ref{fig:hierarchy-fractions} and the parameter-driver plots in Figs.~\ref{fig:driver-ranking}--\ref{fig:alignment-bxsgamma-correlations} can be read as the corrected-rate phenomenology of the final 542-point sample.  The detailed size and interpretation of the relative one-loop shifts are discussed later in Sec.~\ref{subsec:one-loop-results}.

\subsection{Branching-ratio hierarchy}
\label{subsec:br-hierarchy}

The eight flavor-violating channels considered in this work display a pronounced hierarchy in the surviving sample, with the ordering quoted in the rate convention of each channel.  The numerically largest rates are carried by $h\to sb$, $h\to\mu\tau$, $h\to db$, and $t\to ch$; the up-sector Higgs-FCNC mode $h\to uc$, the electron-flavor HFV Higgs modes, and $t\to uh$ are much smaller.

\begin{figure}[!htbp]
  \centering
  \includegraphics[width=0.70\textwidth]{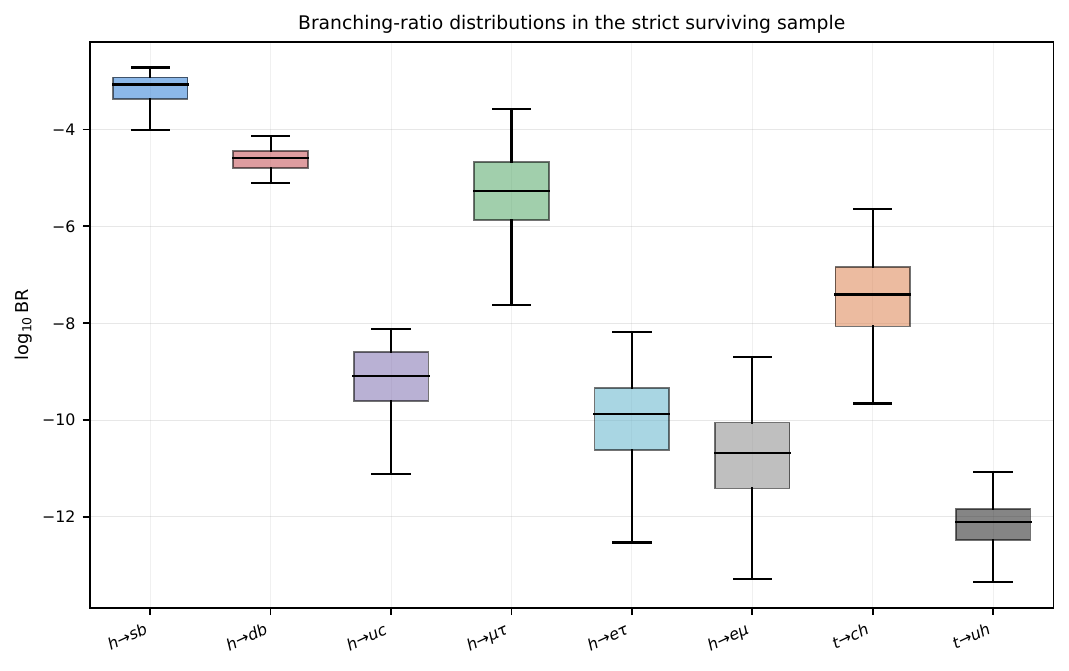}
  \caption{Overview of the flavor-violating rates for the eight channels, shown as $\log_{10}\BR$. Outliers are excluded from the plot.}
  \label{fig:br-overview}
\end{figure}

\FloatBarrier
As shown in Fig.~\ref{fig:br-overview}, the median rates in this convention are approximately $8.43\times10^{-4}$ for $h\to sb$, $2.59\times10^{-5}$ for $h\to db$, $8.04\times10^{-10}$ for $h\to uc$, $5.41\times10^{-6}$ for $h\to\mu\tau$, $3.93\times10^{-8}$ for $t\to ch$, and $7.85\times10^{-13}$ for $t\to uh$. The corresponding maximum branching ratios are $1.91\times10^{-3}$ for $h\to sb$, $7.20\times10^{-5}$ for $h\to db$, $7.57\times10^{-9}$ for $h\to uc$, $2.70\times10^{-4}$ for $h\to\mu\tau$, $6.68\times10^{-9}$ for $h\to e\tau$, $2.01\times10^{-9}$ for $h\to e\mu$, $2.32\times10^{-6}$ for $t\to ch$, and $8.59\times10^{-12}$ for $t\to uh$. The resulting qualitative pattern is that the two down-sector Higgs-FCNC modes satisfy
\[
 \BR(h\to sb)\gg \BR(h\to db),
\]
while the up-sector Higgs-FCNC mode $h\to uc$ is much smaller than the down-sector modes in the final sample. The lepton hierarchy is
\[
 \BR(h\to\mu\tau)\gg \BR(h\to e\tau),\BR(h\to e\mu) .
\]
For the top channels the ordering is \(\BR(t\to ch)\gg\BR(t\to uh)\), with a median ratio \(\BR(t\to ch)/\BR(t\to uh)\simeq 6.46\times10^{4}\).

\begin{figure}[!htbp]
  \centering
  \includegraphics[width=0.96\textwidth]{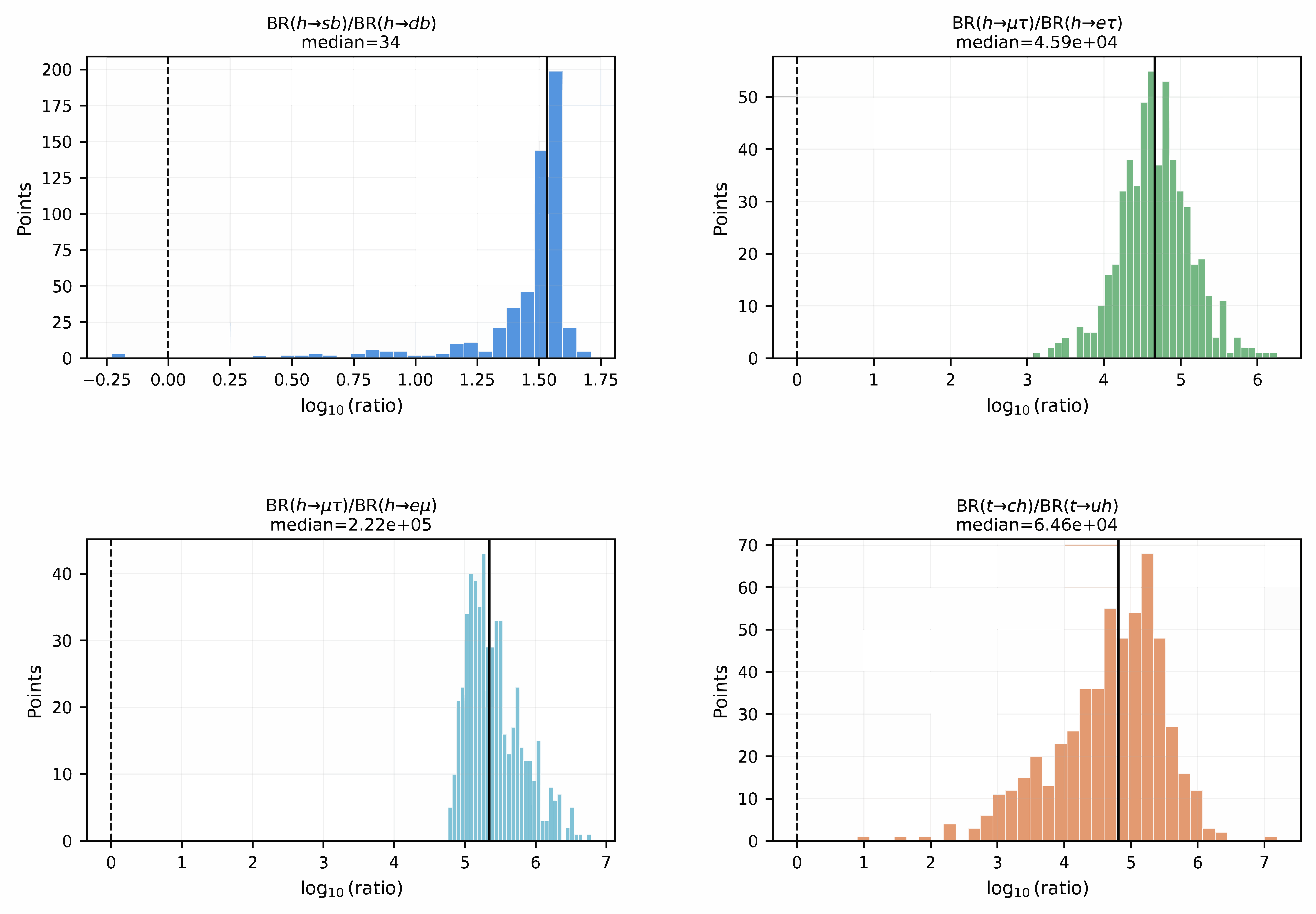}
  \caption{Hierarchy-ratio distributions in the Type 1B F2HDM sample. The dashed black vertical line marks unit ratio, i.e. $\log_{10}(\mathrm{ratio})=0$, while the solid black vertical line marks the median of each distribution.}
  \label{fig:hierarchy-ratios}
\end{figure}

\FloatBarrier
\begin{figure}[!htbp]
  \centering
  \includegraphics[width=0.65\textwidth]{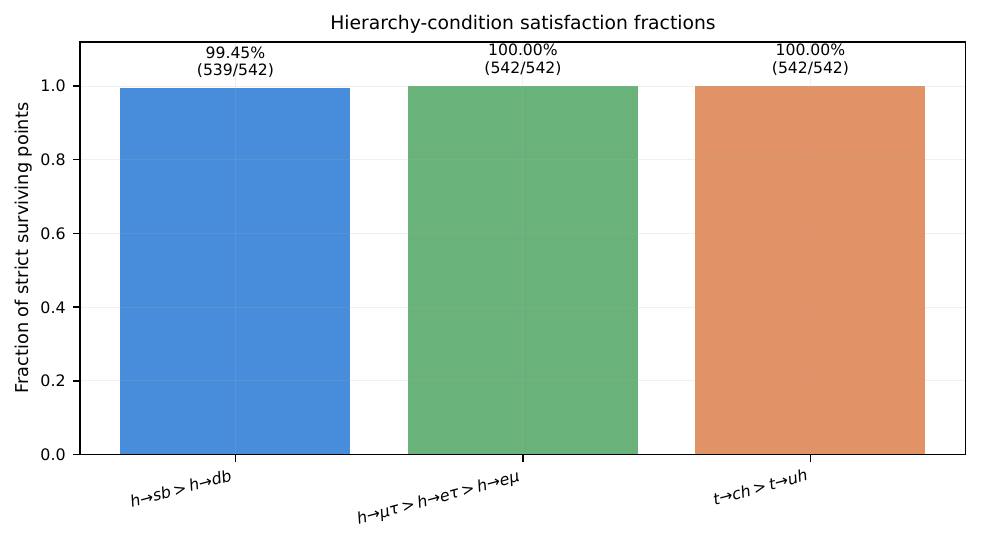}
  \caption{Fractions of points satisfying the down-sector, lepton-sector, and top-sector hierarchy conditions.}
  \label{fig:hierarchy-fractions}
\end{figure}

\FloatBarrier
The hierarchy distributions in Figs.~\ref{fig:hierarchy-ratios} and \ref{fig:hierarchy-fractions} show stable down-sector, lepton, and top patterns across the final sample. The median ratio $\BR(h\to sb)/\BR(h\to db)$ is about $34.0$ after the one-loop rate update, and the down-sector condition holds for $539/542=99.45\%$ of the sample. The lepton and top hierarchy conditions hold for all final points, with median $\BR(t\to ch)/\BR(t\to uh)\simeq 6.46\times10^{4}$. The three points with $\BR_{\mathrm{1L}}(h\to db)>\BR_{\mathrm{1L}}(h\to sb)$ are discussed in Sec.~\ref{subsec:hierarchy-inversion}.

\subsection{Parameter drivers and low-energy correlations}
\label{subsec:drivers}

To identify monotonic parameter dependences we use Spearman's rank coefficient,
\begin{equation}
\rho_S(X,Y)=\mathrm{corr}[\mathrm{rank}(X),\mathrm{rank}(Y)],
\end{equation}
usually with \(Y=\log_{10}\BR\). We use the alignment proxy \(\kappa_V\) defined in Sec.~\ref{subsec:scalar-sector}. We also use the effective texture combination
\begin{equation}
O^{f,ij}_{\rm eff}=\sqrt{\frac{|O^f_{ij}|^2+|O^f_{ji}|^2}{2}},\qquad f=u,d,\ell,
\end{equation}
which is a derived quantity.

\begin{figure}[!htbp]
  \centering
  \includegraphics[width=0.95\textwidth]{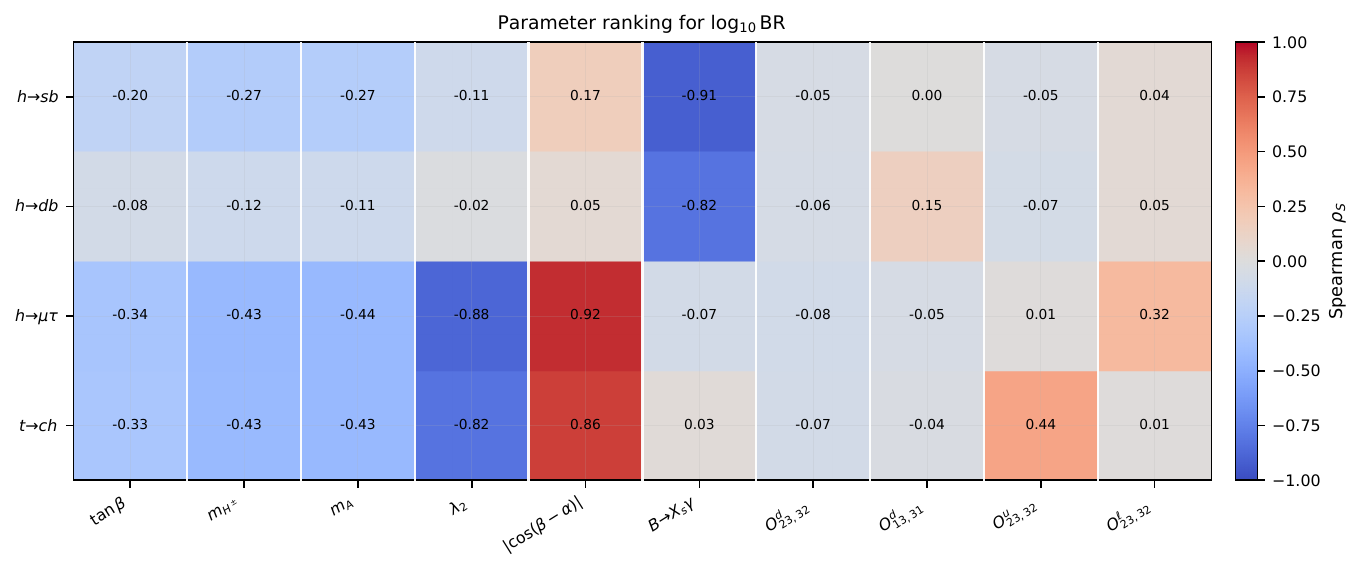}
  \caption{Spearman-rank driver analysis for the main flavor-violating rates.}
  \label{fig:driver-ranking}
\end{figure}

\FloatBarrier
The driver ranking in Fig.~\ref{fig:driver-ranking} shows that $h\to sb$ and $h\to db$ are shaped mainly by the constrained down-sector structure and by low-energy flavor selections, with no dominant single $O(1)$ texture entry. Numerically, $\rho_S(O^{d,23}_{\rm eff})=-0.055$ for $h\to sb$ and $\rho_S(O^{d,13}_{\rm eff})=0.151$ for $h\to db$, while the strongest monotonic correlation in the two down-sector rates is with $\BR(B\to X_s\gamma)$, $\rho_S=-0.909$ and $-0.819$, respectively. The $h\to\mu\tau$ rate remains strongly alignment controlled, with $\rho_S(|1-\kappa_V|)=0.916$, and the $t\to ch$ rate also has a positive alignment correlation, $\rho_S(|1-\kappa_V|)=0.862$.

\begin{figure}[!htbp]
  \centering
  \includegraphics[width=0.92\textwidth]{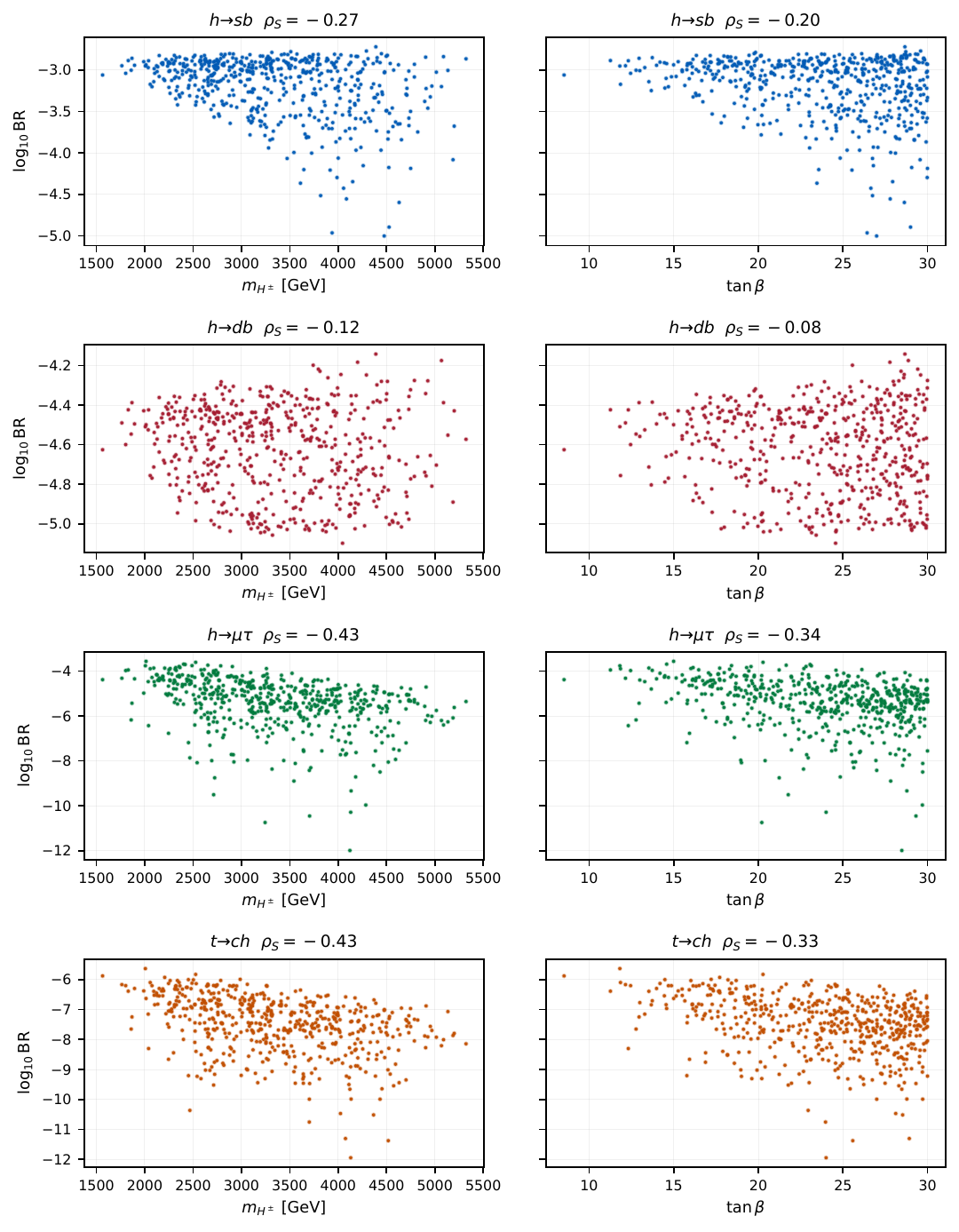}
  \caption{Dependence of the main flavor-violating rates on $m_{H^\pm}$ and $\tan\beta$.}
  \label{fig:mhpm-tanbeta-dependence}
\end{figure}

\begin{figure}[!htbp]
  \centering
  \includegraphics[width=0.95\textwidth]{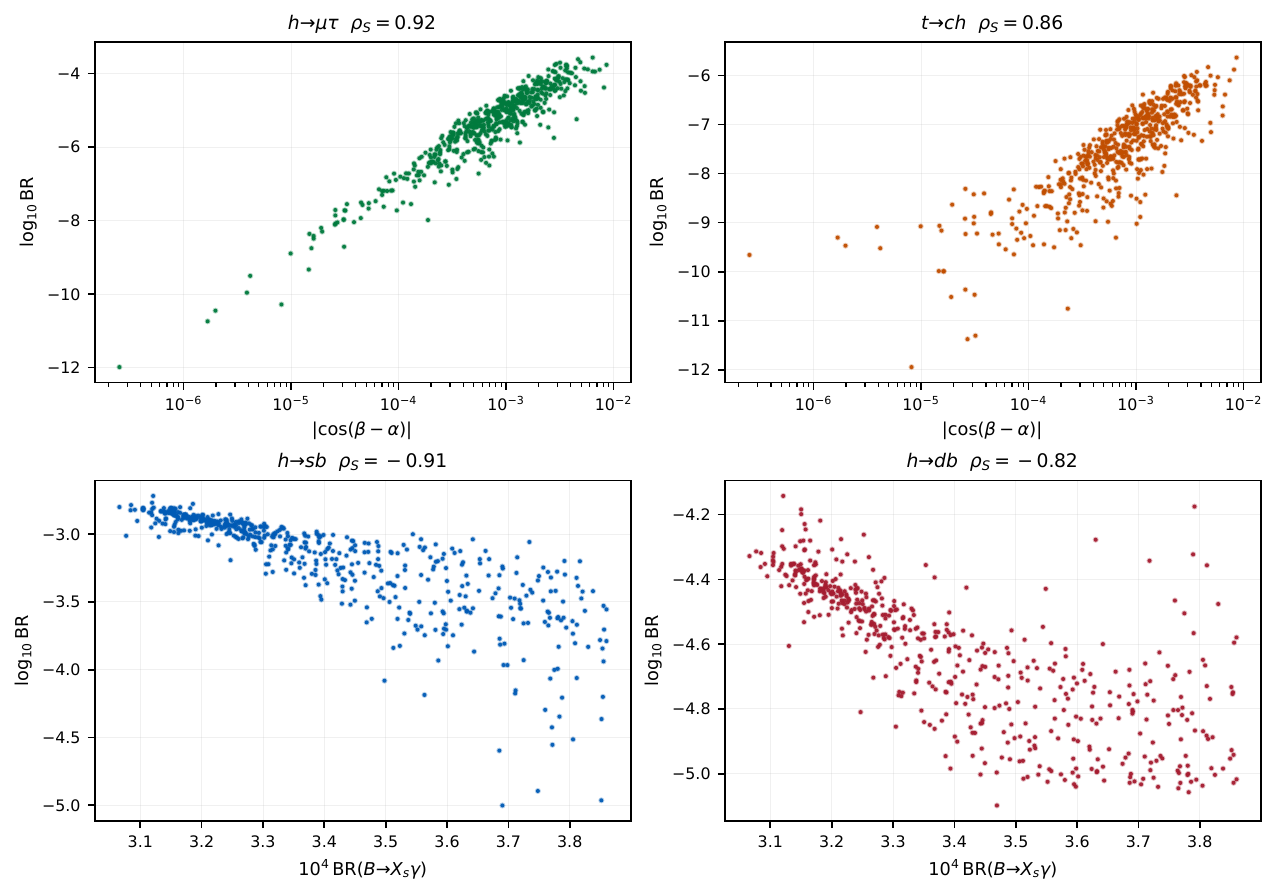}
  \caption{Alignment and $B\to X_s\gamma$ correlations for the main flavor-violating rates.}
  \label{fig:alignment-bxsgamma-correlations}
\end{figure}

\FloatBarrier
Figure~\ref{fig:mhpm-tanbeta-dependence} shows that the down-sector Higgs-FCNC rates in the final sample retain only mild direct monotonic dependence on $m_{H^\pm}$ and $\tan\beta$ once the stronger flavor selections are imposed; for example $\rho_S(\tan\beta,\log_{10}\BR)=-0.200$ for $h\to sb$ and $-0.075$ for $h\to db$. The $h\to\mu\tau$ and $t\to ch$ rates are more visibly alignment controlled, with additional negative correlations with $\lambda_2$ in the driver ranking. The alignment panels in Fig.~\ref{fig:alignment-bxsgamma-correlations} show that $h\to\mu\tau$ remains strongly alignment-controlled, while the $t\to ch$ rate has a positive alignment correlation and a compressed absolute range. The same figure also shows the strong $B\to X_s\gamma$ correlation in the down-sector channels.

\subsection{One-loop vertex effects}
\label{subsec:one-loop-results}

Having fixed the corrected-rate convention above, we now quantify the size of the vertex effects in Higgs-mediated FCNC decays and rare top decays.  The relative quantities in Table~\ref{tab:delta-summary} are computed with the triangle categories shown in Figs.~\ref{fig:one-loop-triangle-hdd}--\ref{fig:one-loop-triangle-tqh}.

\begin{table}[!htbp]
\centering
\small
\resizebox{0.92\textwidth}{!}{%
\begin{tabular}{|l|r|r|r|r|r|}
\hline
Channel & min & 5\% & median & 95\% & max \\
\hline
$t\to ch$ & $-0.956$ & $-0.248$ & $-0.0927$ & $3.07$ & $3.06\times 10^{4}$ \\
$t\to uh$ & $-0.91$ & $-0.0931$ & $1.11$ & $485$ & $2.74\times 10^{7}$ \\
$h\to uc$ & $-0.716$ & $49.7$ & $4.09\times 10^{3}$ & $2.14\times 10^{6}$ & $4.25\times 10^{10}$ \\
$h\to sb$ & $-0.993$ & $-0.906$ & $-0.444$ & $-0.0473$ & $0.252$ \\
$h\to db$ & $-0.828$ & $-0.78$ & $-0.422$ & $0.0485$ & $0.624$ \\
\hline
\end{tabular}%
}
\caption{One-loop relative-correction quantiles for the quark and top FCNC channels in the strict surviving sample.}
\label{tab:delta-summary}

\end{table}

For $t\to ch$, very large positive relative corrections occur in the one-loop
rate comparison.  These large ratios are primarily denominator effects caused by
very small tree-level rates near the alignment region; the corresponding absolute
physical branching ratios remain small.
For this channel the relative correction can be written as
\begin{equation*}
\delta_{\mathrm{1L}}
=
\frac{\left|\mathcal M_{\rm tree}+\mathcal M_{\rm loop}\right|^2
-\left|\mathcal M_{\rm tree}\right|^2}
{\left|\mathcal M_{\rm tree}\right|^2}
=
2\,{\rm Re}\!\left(\frac{\mathcal M_{\rm loop}}{\mathcal M_{\rm tree}}\right)
+\left|\frac{\mathcal M_{\rm loop}}{\mathcal M_{\rm tree}}\right|^2 .
\end{equation*}
Near the alignment region, the tree-level \(tch\) coupling is strongly
suppressed, so \(\BR_{\rm tree}(t\to ch)\) becomes very small.

\begin{figure}[!htbp]
  \centering
  \includegraphics[width=0.95\textwidth]{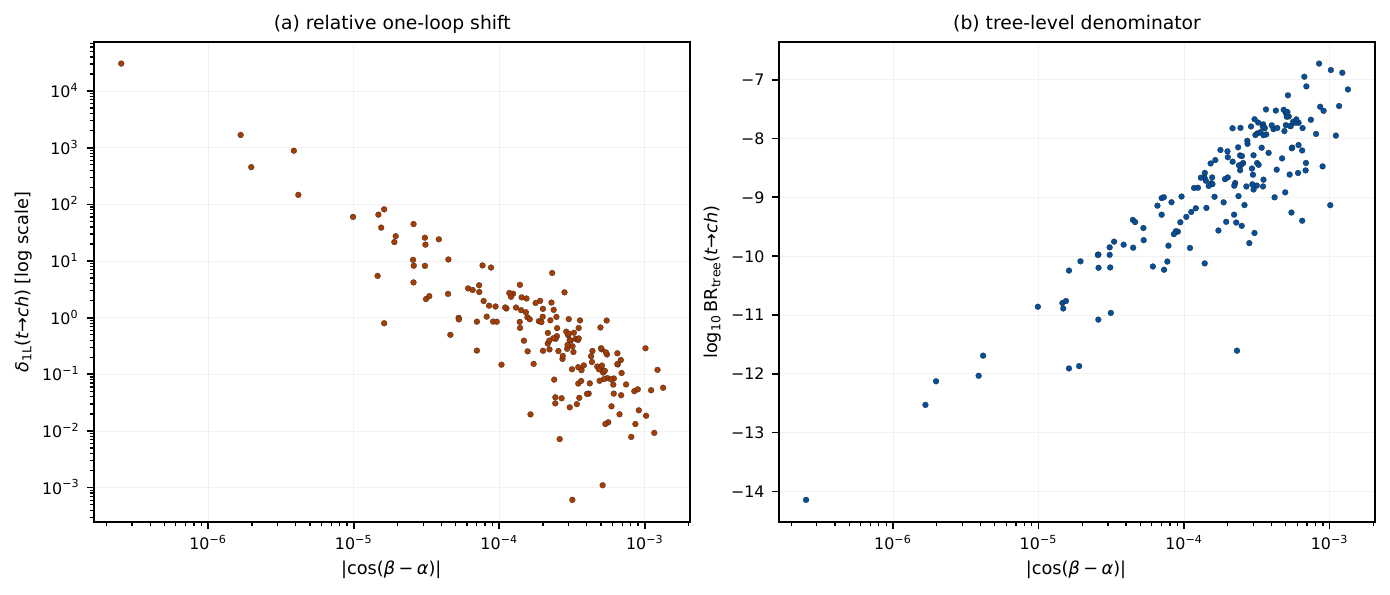}
  \caption{Alignment dependence in the $t\to ch$ channel: (a) $\delta_{\mathrm{1L}}$ on a logarithmic scale and (b) $\log_{10}\BR_{\rm tree}$ as functions of $|\cos(\beta-\alpha)|$.}
  \label{fig:tch-one-loop-correction}
\end{figure}

Figure~\ref{fig:tch-one-loop-correction} illustrates this denominator effect in
more detail.  In panel (a), the largest values of $\delta_{\mathrm{1L}}$ appear
close to the alignment region.  Panel (b) shows that the same region contains
very small tree-level $t\to ch$ branching ratios.  The corresponding
one-loop-updated branching ratio is shown in
Fig.~\ref{fig:alignment-bxsgamma-correlations}(b), where the absolute
$t\to ch$ rate remains small and compressed.  The large values of
\(\delta_{\mathrm{1L}}\) should therefore be interpreted as relative ratios in
tree-suppressed regions, not as evidence for a parametrically large absolute
one-loop-induced \(t\to ch\) branching ratio.
For $h\to sb$ and
$h\to db$, the median one-loop relative corrections are negative,
\(\delta_{\mathrm{1L}}\simeq -0.444\) and \(-0.422\), respectively,
indicating a net suppression in the down-sector channels.

\subsection{Tree-level stability and CKM-down origin}
\label{subsec:tree-stability}

A striking feature of the surviving sample is the different width of the tree-level branching-ratio distributions. The tree-level quantiles in Table~\ref{tab:tree-br-quantiles} show that $h\to sb$ and $h\to db$ are locked into narrow bands, while representative up-sector Higgs and top FCNC rates span several orders of magnitude.

\begin{table}[!htbp]
\centering
\small
\resizebox{0.92\textwidth}{!}{%
\begin{tabular}{|l|c|c|c|c|c|}
\hline
Channel & Minimum & 5th percentile & Median & 95th percentile & Maximum \\
\hline
$h\to sb$ & $1.37\times 10^{-3}$ & $1.46\times 10^{-3}$ & $1.54\times 10^{-3}$ & $1.57\times 10^{-3}$ & $1.57\times 10^{-3}$ \\
$h\to db$ & $3.81\times 10^{-5}$ & $4.14\times 10^{-5}$ & $4.49\times 10^{-5}$ & $4.64\times 10^{-5}$ & $4.66\times 10^{-5}$ \\
$t\to ch$ & $7.17\times 10^{-15}$ & $1.08\times 10^{-10}$ & $4.65\times 10^{-8}$ & $7.17\times 10^{-7}$ & $2.71\times 10^{-6}$ \\
$h\to \mu\tau$ & $1.04\times 10^{-12}$ & $1.91\times 10^{-8}$ & $5.27\times 10^{-6}$ & $8.26\times 10^{-5}$ & $2.65\times 10^{-4}$ \\
\hline
\end{tabular}%
}
\caption{Tree-level branching-ratio quantiles for the main channels in the strict surviving sample.}
\label{tab:tree-br-quantiles}

\end{table}

The down-sector central 5--95\% intervals are $[1.46,1.57]\times10^{-3}$ for $h\to sb$ and $[4.14,4.64]\times10^{-5}$ for $h\to db$. In contrast, the corresponding intervals are $1.08\times10^{-10}$ to $7.17\times10^{-7}$ for $t\to ch$ and $3.51\times10^{-16}$ to $5.35\times10^{-12}$ for $h\to uc$.

The stability can be traced directly to the mass-basis reconstruction in Sec.~\ref{subsec:yukawa-lagrangian}. In the down-sector matrix of Eq.~\eqref{eq:mprime-down}, the entries relevant for \(h\to sb\) contain
\begin{equation}
(M'_d)_{23}=-V_{ts}^*m_b\left(1+O^d_{23}\frac{m_s}{m_b}\right),\qquad
(M'_d)_{32}=O^d_{32}m_s .
\end{equation}
The dependence on \(O^d_{23}\) is suppressed by \(m_s/m_b\), so the leading structure is the fixed CKM-weighted term \(-V_{ts}^*m_b\); the opposite-chirality-related entry is only \(m_s\)-sized. Similarly, for \(h\to db\),
\begin{equation}
(M'_d)_{13}=-V_{td}^*m_b\left(1+O^d_{13}\frac{m_s}{m_b}\right),\qquad
(M'_d)_{31}=O^d_{31}m_d .
\end{equation}
The leading fixed term \(-V_{td}^*m_b\) dominates over the scanned \(O^d\)-dependent pieces. This explains why the down-sector tree-level branching ratios are only weakly sensitive to the scanned \(O^d\) factors and form narrow bands. Their remaining variation is tied to the fixed CKM factors, mass reconstruction, Higgs mixing/alignment, and the surviving flavor and electroweak constraints.

This behavior contrasts with \(t\to ch\), \(h\to uc\), and the lepton-sector \(h\to\mu\tau\) rate. In Eq.~\eqref{eq:mprime-up}, the relevant up-sector off-diagonal entries are controlled by the scanned \(O^u\) texture factors and by alignment-dependent Higgs mixing; the fixed CKM-weighted \(m_b\) leading term is a special feature of the down-sector entries. In Eq.~\eqref{eq:mprime-lepton}, the relevant lepton-sector entries are controlled by \(O^\ell_{23}m_\mu\) and \(O^\ell_{32}m_\mu\). Consequently, the non-down-sector flavor-violating rates remain more sensitive to scanned texture entries, alignment, and accidental cancellations among Higgs-doublet contributions.

\begin{figure}[!htbp]
  \centering
  \includegraphics[width=0.95\textwidth]{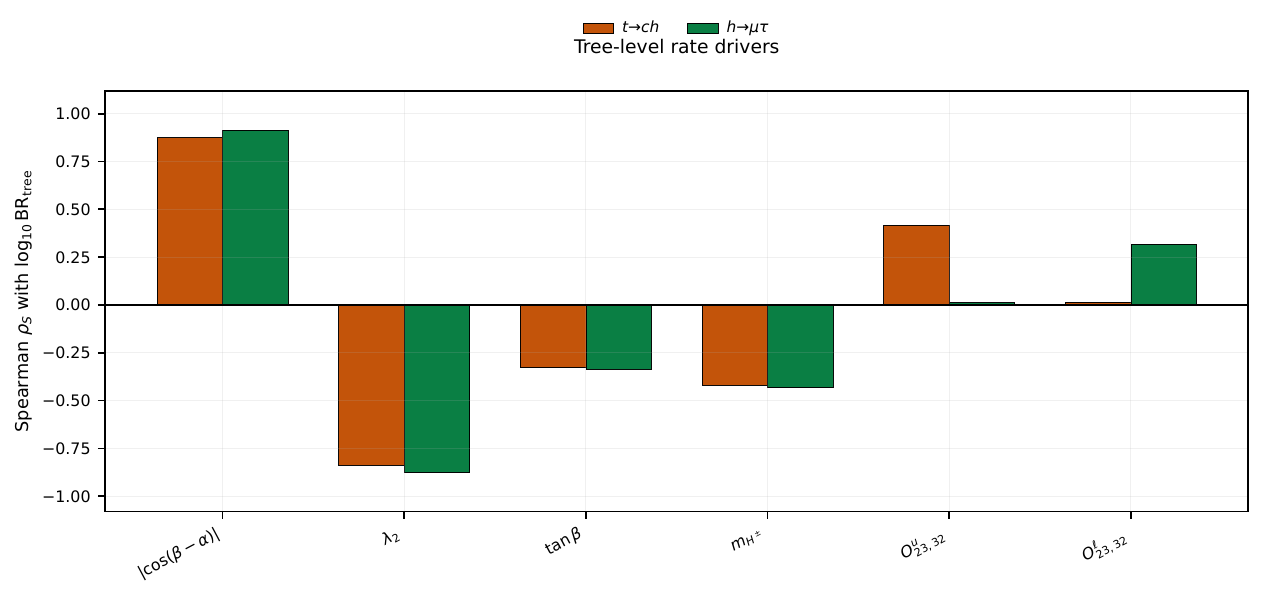}
  \caption{Spearman correlations for representative tree-level Higgs and top FCNC branching ratios in the final sample.}
  \label{fig:tree-driver-bars}
\end{figure}

\begin{figure}[!htbp]
  \centering
  \includegraphics[width=0.95\textwidth]{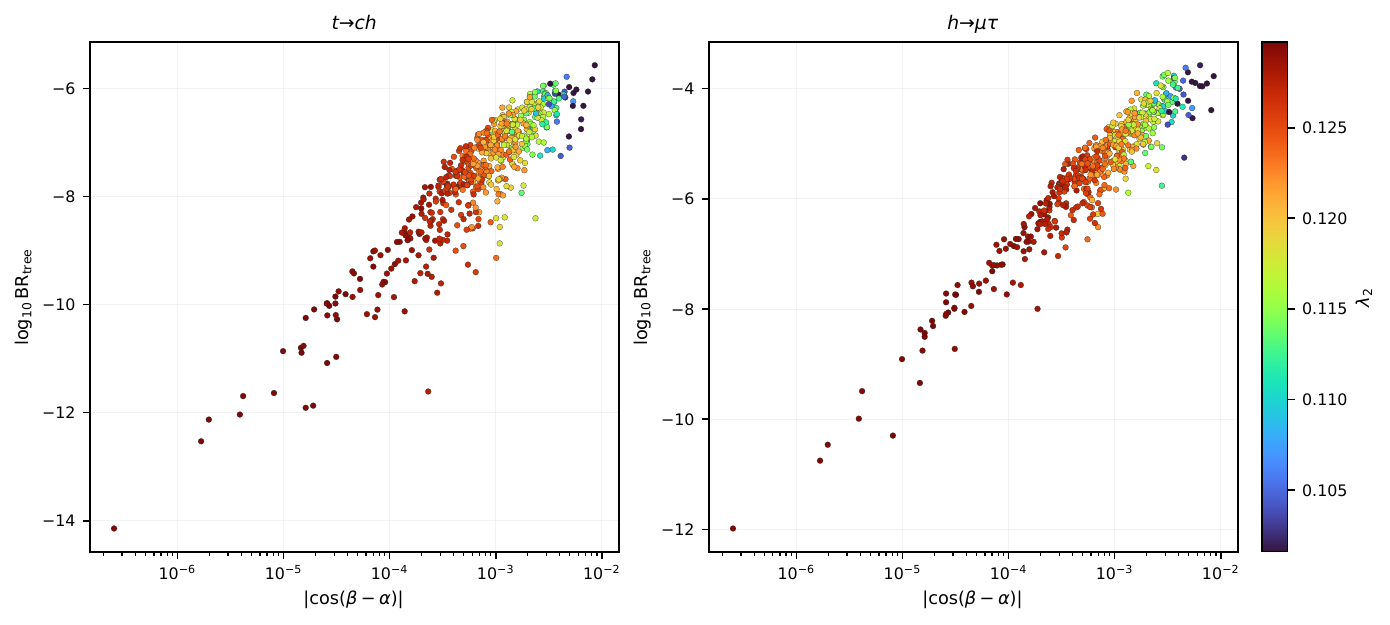}
  \caption{Correlation of $\lambda_2$ and the alignment proxy with the loop-level $h\to\mu\tau$ rate and representative tree-level FCNC rates.}
  \label{fig:lambda2-alignment-tree-rates}
\end{figure}

The broad non-down-sector distributions are visible in the driver plot of Fig.~\ref{fig:tree-driver-bars} and the $\lambda_2$--alignment scatter plot of Fig.~\ref{fig:lambda2-alignment-tree-rates}. These channels reflect their direct sensitivity to the scanned up- and lepton-sector textures, the alignment limit, and accidental cancellations among Higgs-doublet contributions.

\subsection{Rare hierarchy inversion in \texorpdfstring{$h\to db$ versus $h\to sb$}{h to db versus h to sb}}
\label{subsec:hierarchy-inversion}

A small group of three points satisfies
\(\BR_{\mathrm{1L}}(h\to db)>\BR_{\mathrm{1L}}(h\to sb)\) after the one-loop rate is applied.  All three inversion points retain
the tree-level ordering
\(\BR_{\rm tree}(h\to sb)>\BR_{\rm tree}(h\to db)\), as summarized in
Table~\ref{tab:hierarchy-inversion}.  We define
\begin{equation}
K_{sb}=\frac{\BR_{\mathrm{1L}}(h\to sb)}{\BR_{\rm tree}(h\to sb)},\qquad
K_{db}=\frac{\BR_{\mathrm{1L}}(h\to db)}{\BR_{\rm tree}(h\to db)} .
\end{equation}

\begin{table}[!htbp]
\centering
\small
\resizebox{0.86\textwidth}{!}{%
\begin{tabular}{|l|c|}
\hline
Quantity & Inversion points \\
\hline
Number of inversion points & 3 \\
Inversion fraction & 0.55\% \\
Median tree ratio $\BR_{\rm tree}(h\to sb)/\BR_{\rm tree}(h\to db)$ & $34.2$ \\
Median one-loop ratio $\BR_{\mathrm{1L}}(h\to sb)/\BR_{\mathrm{1L}}(h\to db)$ & $0.592$ \\
Median $K_{sb}=\BR_{\mathrm{1L}}(h\to sb)/\BR_{\rm tree}(h\to sb)$ & $7.08\times 10^{-3}$ \\
Median $K_{db}=\BR_{\mathrm{1L}}(h\to db)/\BR_{\rm tree}(h\to db)$ & $0.412$ \\
Median $K_{db}/K_{sb}$ & $57.5$ \\
All inversion points keep tree-level $h\to sb>h\to db$ & yes \\
\hline
\end{tabular}%
}
\caption{Summary of the rare $h\to db>h\to sb$ hierarchy-inversion points after applying one-loop rates in the strict surviving sample.}
\label{tab:hierarchy-inversion}

\end{table}

The inversion appears as a correction-factor inversion and is traced to
destructive tree--loop interference in the down-sector channels.  In the final sample, the inverted points have $K_{sb}\simeq (6.5\text{--}8.2)\times 10^{-3}$ and $K_{db}\simeq 0.34\text{--}0.47$, with median values $K_{sb}=7.1\times 10^{-3}$ and $K_{db}=0.41$.  Thus \(K_{db}/K_{sb}\) is large enough to overcome the
underlying tree-level hierarchy even though the tree-level branching ratios keep
\(h\to sb\) above \(h\to db\).  Schematically,
\begin{equation}
K_q =
\frac{\left|\mathcal M_{\rm tree}^{\,q}+\mathcal M_{\rm loop}^{\,q}\right|^2}
{\left|\mathcal M_{\rm tree}^{\,q}\right|^2}
=1+\delta_{\rm int}^{q}+R_{\rm loop}^{q},
\qquad q=sb,db,
\end{equation}
with
\begin{equation}
\begin{aligned}
&\delta_{\rm int}^{q}
=
\frac{2\,{\rm Re}\!\left[
\left(\mathcal M_{\rm tree}^{\,q}\right)^*
\mathcal M_{\rm loop}^{\,q}\right]}
{\left|\mathcal M_{\rm tree}^{\,q}\right|^2},\\
&R_{\rm loop}^{q}
=
\frac{\left|\mathcal M_{\rm loop}^{\,q}\right|^2}
{\left|\mathcal M_{\rm tree}^{\,q}\right|^2},\\
&C_{\rm int}^{q}
\equiv
\frac{{\rm Re}\!\left[
\left(\mathcal M_{\rm tree}^{\,q}\right)^*
\mathcal M_{\rm loop}^{\,q}\right]}
{\left|\mathcal M_{\rm tree}^{\,q}\right|\,
 \left|\mathcal M_{\rm loop}^{\,q}\right|},
\qquad q=sb,db .
\end{aligned}
\end{equation}
The small \(K_{sb}\) values therefore indicate a near cancellation in the total
\(h\to sb\) amplitude, while the corresponding \(h\to db\) rates are suppressed
less severely.

\begin{figure}[!htbp]
  \centering
  \includegraphics[width=0.92\textwidth]{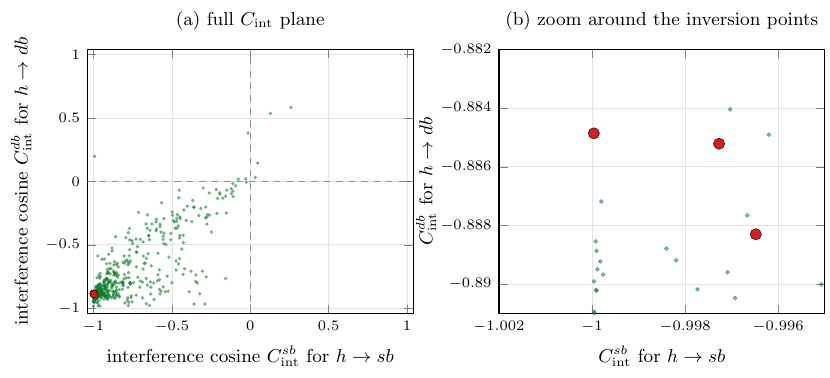}
    \caption{Interference-cosine projection of the rare hierarchy inversion.  Panel (a) shows the full \(C_{\mathrm{int}}^{sb}\)--\(C_{\mathrm{int}}^{db}\) plane, and panel (b) zooms into the inversion region.  Green points show non-inverted points, while red points show the three points with \(\BR_{\mathrm{1L}}(h\to db)>\BR_{\mathrm{1L}}(h\to sb)\), after summing the charge-conjugate final states.}
\label{fig:cint-inversion-wide}
\end{figure}

Figure~\ref{fig:cint-inversion-wide} shows that the three inverted points lie in
the region with strongly negative \(C_{\mathrm{int}}^{sb}\), while
\(C_{\mathrm{int}}^{db}\) is less extreme.  The zoom panel resolves the three
surviving inverted points individually.  The presence of these three inversion
points shows that the destructive-interference mechanism survives the stringent
meson-mixing constraints.

\subsection{Lepton-sector HFV tree-level and one-loop comparison}
\label{subsec:lepton-tree-loop-comparison}

\begin{figure}[!htbp]
  \centering
  \begin{subfigure}{0.46\textwidth}
    \centering
    \includegraphics[width=\linewidth]{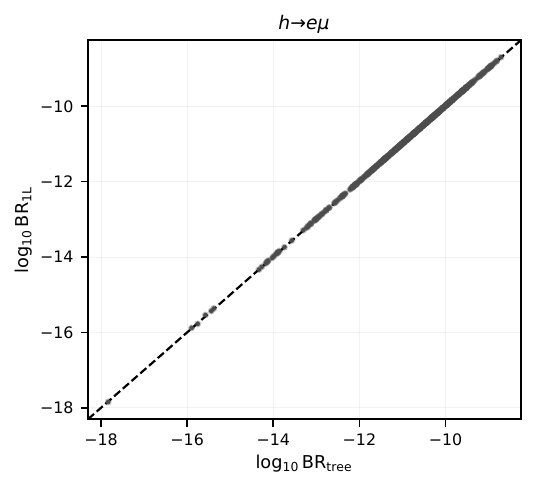}
  \end{subfigure}\hfill
  \begin{subfigure}{0.46\textwidth}
    \centering
    \includegraphics[width=\linewidth]{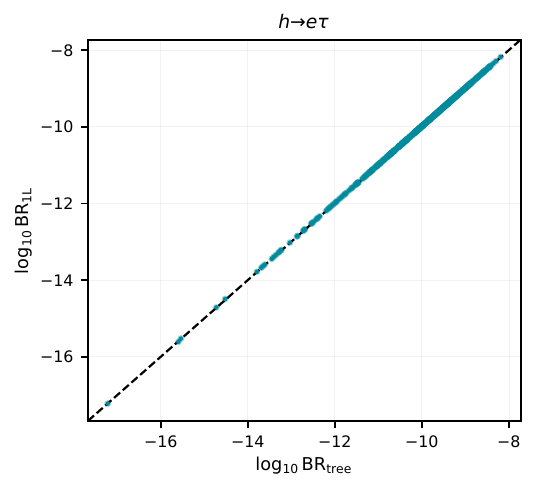}
  \end{subfigure}

  \vspace{0.8ex}
  \begin{subfigure}{0.46\textwidth}
    \centering
    \includegraphics[width=\linewidth]{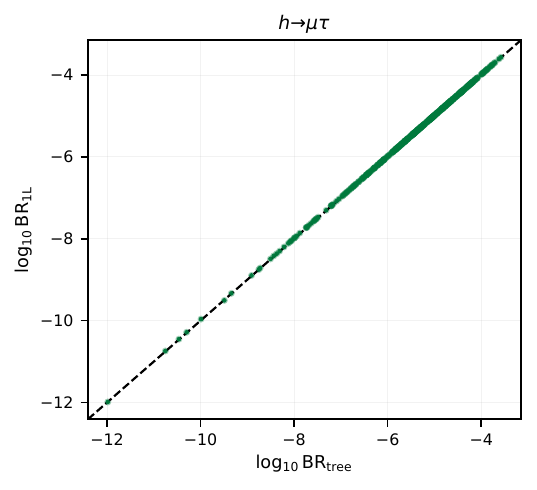}
  \end{subfigure}
  \caption{Comparison of the tree-level and SPheno one-loop branching ratios for the lepton-sector HFV channels in the 542-point final sample. The dashed diagonal line in each panel denotes \(\mathrm{BR}_{1L}=\mathrm{BR}_{\mathrm{tree}}\).}
  \label{fig:hll-tree-vs-oneloop}
\end{figure}

In the final sample, Fig.~\ref{fig:hll-tree-vs-oneloop} compares the tree-level and SPheno one-loop branching ratios for the three lepton-sector HFV Higgs channels. The points lie close to the equal-rate line, showing that the one-loop corrections are generally small over the surviving parameter space.

\FloatBarrier

\subsection{Experimental prospects and interpretation}
\label{subsec:experimental-prospects}

The hierarchy found in Sec.~\ref{subsec:br-hierarchy} is a model hierarchy in the surviving Type 1B parameter sample; LHC observability requires separate detector-level information. The channel \(h\to sb\) can reach the \(10^{-3}\) level in branching ratio, but its experimental reconstruction is challenging because the final state is embedded in ordinary hadronic Higgs decays and requires flavor-sensitive strange- and bottom-jet information. A direct collider interpretation would therefore need a dedicated analysis with realistic \(b\)-tagging, possible strange-tagging or strange-enriched categories, and detector-level efficiencies.

\begin{table}[!htbp]
\centering
\caption{Current direct experimental upper limits or direct-limit status for the flavor-violating channels considered in this work.}
\label{tab:direct-exp-limits}
\begin{ruledtabular}
\begin{tabular}{l p{0.58\textwidth} l}
Channel & Direct bound/status & Source \\
\(h\to e\mu\) & \(\BR<4.4\times10^{-5}\) at \(95\%\) C.L. & CMS~\cite{CMS:2023xpx} \\
\(h\to e\tau\) & \(\BR<2.0\times10^{-3}\) at \(95\%\) C.L. & ATLAS~\cite{ATLAS:2023mvd} \\
\(h\to\mu\tau\) & \(\BR<1.5\times10^{-3}\) at \(95\%\) C.L. & CMS~\cite{CMS:2021rsq} \\
\(t\to uh\) & \(\BR<2.6\times10^{-4}\) at \(95\%\) C.L. & ATLAS~\cite{ATLAS:2024topHml} \\
\(t\to ch\) & \(\BR<3.4\times10^{-4}\) at \(95\%\) C.L. & ATLAS~\cite{ATLAS:2024topHml} \\
\(h\to sb,\ h\to db,\ h\to uc\) & No dedicated process-specific direct upper limit is currently available; present constraints on quark-flavor Higgs couplings are indirect or analysis dependent. & \cite{Harnik:2012pb,Herrero-Garcia:2019mcy} \\
\end{tabular}
\end{ruledtabular}
\end{table}

The lepton-sector HFV mode \(h\to\mu\tau\) is experimentally cleaner. The limits in Table~\ref{tab:direct-exp-limits} show \(10^{-3}\)-level sensitivity in \(h\to e\tau\) and \(h\to\mu\tau\), and a stronger \(10^{-5}\)-level bound in \(h\to e\mu\). The values obtained in the surviving sample are below current bounds but can lie in a range relevant for improved Run-3 and high-luminosity LHC analyses. By contrast, \(h\to e\tau\) and \(h\to e\mu\) are typically much smaller in this Type 1B realization.

For top-Higgs FCNCs, current direct limits reach the \(10^{-4}\) level. Both the tree-level texture hierarchy and the loop-corrected rates emphasize \(t\to ch\) over \(t\to uh\) in the surviving sample. Future reinterpretations should combine the model-specific flavor structure with realistic top reconstruction, Higgs decay modes, flavor tagging, and correlations with the low-energy constraints. Thus the most promising phenomenological targets depend on both rate size and experimental accessibility: \(h\to sb\) is numerically large but experimentally difficult, \(h\to\mu\tau\) is cleaner, and \(t\to ch\) provides the leading top-flavor probe in this numerical analysis.

\section{Conclusions}
\label{sec:conclusions}

We have studied flavor-violating Higgs and top decays in the Type 1B flavorful two-Higgs-doublet model with a twist. Generation-dependent doublet assignments make the Yukawa sector neither natural-flavor-conserving nor aligned, so tree-level scalar FCNCs arise. They are organized by the approximate $U(2)^5$ texture and, in our implementation, by a CKM matrix generated in the down sector. Unlike a generic 2HDM with tree-level FCNCs, where off-diagonal Yukawa couplings are often treated as more freely parameterized, Type 1B ties them to these assignments and the CKM-down mass reconstruction, yielding correlated flavor hierarchies.

After imposing Higgs, electroweak, meson-mixing, $B\to X_s\gamma$, HFV, and CLFV constraints (including $\mu$--$e$ conversion), the surviving sample exhibits $\BR(h\to sb)\gg\BR(h\to db)$, $\BR(h\to\mu\tau)\gg\BR(h\to e\tau),\BR(h\to e\mu)$, and $\BR(t\to ch)\gg\BR(t\to uh)$. The leading channels are $h\to sb$, $t\to ch$, $h\to db$, and $h\to\mu\tau$. The tree-level rates for $h\to sb$ and $h\to db$ are narrowly distributed and CKM-controlled by the fixed $-V_{ts}^*m_b$ and $-V_{td}^*m_b$ terms, whereas $t\to ch$ and $h\to\mu\tau$ vary by orders of magnitude because of texture and alignment sensitivity. Approximate first--second-generation protection suppresses $h\to uc$ and $h\to e\mu$. Radiative and three-body CLFV observables remain below current limits and are not used as additional sample-reducing cuts.

We also included one-loop vertex effects, including the loop-amplitude-squared contribution to the rates. Large relative corrections in $t\to ch$ occur when alignment suppresses the tree-level coupling, although the absolute branching ratio remains small. Rare points with $\BR_{\mathrm{1L}}(h\to db)>\BR_{\mathrm{1L}}(h\to sb)$ arise from strong destructive tree--loop interference that suppresses $h\to sb$ more strongly than $h\to db$. The overall pattern nevertheless remains $h\to sb$ as the largest quark-HFV mode, $h\to\mu\tau$ as the cleanest lepton-HFV mode, and $t\to ch$ as the leading top-flavor probe.

\begin{acknowledgments}
This work was supported by the National Natural Science Foundation
of China (NNSFC) under Grants No. 12075074, 12235008, 11535002, and 11705045; the
Natural Science Foundation for Distinguished Young Scholars of Hebei Province under Grant
No. A2022201017; the Youth Top-notch Talent Support Program of Hebei Province; and
the Midwest Universities Comprehensive Strength Promotion Project.
\end{acknowledgments}

\appendix

\section{Representative \texorpdfstring{$K^0$--$\bar K^0$}{K0-K0bar} mixing contributions}
\label{app:k0-mixing-representative}

\begin{figure*}[t]
  \centering
  \includegraphics[width=0.93\textwidth]{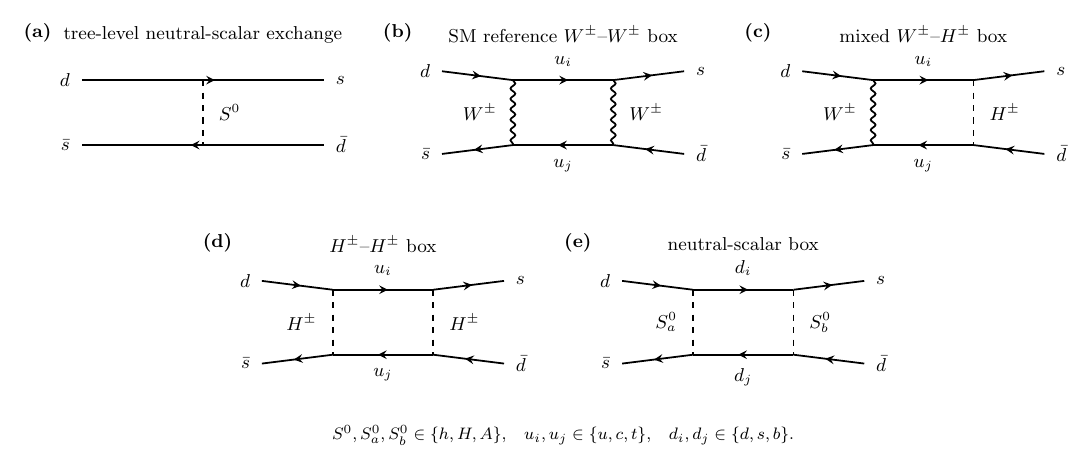}
  \caption{Representative contributions to $K^0$--$\bar K^0$ mixing entering the SPheno/FlavorKit evaluation: (a) tree-level neutral-scalar exchange, (b) SM reference $W^\pm$--$W^\pm$ box, (c) mixed $W^\pm$--$H^\pm$ boxes, (d) $H^\pm$--$H^\pm$ boxes, and (e) neutral-scalar boxes. The mixed $W^\pm$--$H^\pm$ panel represents both boson orderings, and sums over internal flavors are implicit. Additional Goldstone-boson and neutral-gauge contributions present in the generated Wilson coefficients are not displayed separately. The diagrams are representative rather than an exhaustive graphical enumeration.}
  \label{fig:k0-mixing-representative}
\end{figure*}

The meson-mixing observables are evaluated using the SARAH-generated
SPheno/FlavorKit routines~\cite{Porod:2014xia}. Representative contributions to
\(K^0\)--\(\bar K^0\) mixing are shown in
Fig.~\ref{fig:k0-mixing-representative}.

\clearpage

\clearpage

\bibliographystyle{apsrev4-2}
\bibliography{REFERENCES_TYPE1B_F2HDM_FULL_DRAFT}

\end{document}